\documentclass[longauth]{aa}   

\usepackage{graphicx,rotating,epsfig}
\usepackage{txfonts,times,latexsym,amssymb}
\usepackage{natbib}
\usepackage{longtable,lscape,multirow,supertabular}
\usepackage{xcolor}
\usepackage[normalem]{ulem}

\usepackage{tikz} 

\def\kms{km\,s$^{-1}$}
\def\Ebv{\ensuremath{{E}{\mathrm (B-V)}}}
\def\fHtwo{$f_{\text{H}_2}$}

\begin{document}
\title{The ESO Diffuse Interstellar Bands Large Exploration Survey} 
\subtitle{EDIBLES I. Project description, survey sample and quality assessment}

\author{
Nick\,L.\,J.~Cox\inst{1,2} \and
Jan~Cami\inst{3,4} \and
Amin~Farhang\inst{5} \and
Jonathan~Smoker\inst{6} \and
Ana~Monreal-Ibero\inst{7,8,9} \and
Rosine~Lallement\inst{7} \and 
Peter\,J.~Sarre\inst{10} \and
Charlotte\,C.\,M.~Marshall\inst{10} \and
Keith\,T.~Smith\inst{11,12} \and
Christopher\,J.~Evans\inst{13} \and
Pierre~Royer\inst{14} \and
Harold~Linnartz\inst{15} \and
Martin\,A.~Cordiner\inst{16,17} \and
Christine~Joblin\inst{1,2} \and
Jacco\,Th.~van~Loon\inst{18} \and
Bernard\,H.~Foing\inst{19} \and
Neil\,H.~Bhatt\inst{3} \and
Emeric~Bron\inst{20} \and
Meriem~Elyajouri\inst{7} \and
Alex~de Koter\inst{21,14} \and
Pascale~Ehrenfreund\inst{22} \and
Atefeh~Javadi\inst{5} \and
Lex~Kaper\inst{21} \and
Habib\,G.\,~Khosroshadi\inst{5} \and
Mike~Laverick\inst{14} \and
Franck~Le~Petit\inst{23} \and
Giacomo~Mulas\inst{24} \and
Evelyne~Roueff\inst{23} \and
Farid~Salama\inst{25} \and
Marco~Spaans\inst{26} 
}

\institute{
Universit\'e de Toulouse, UPS-OMP, IRAP, 31028, Toulouse, France
\and 
CNRS, IRAP, 9 Av. colonel Roche, BP 44346, F-31028 Toulouse, France
\and
Department of Physics and Astronomy, The University of Western Ontario, London, ON N6A 3K7, Canada
\and
SETI Institute, 189 Bernardo Avenue, Suite 100, Mountain View, CA 94043, USA
\and 
School of Astronomy, Institute for Research in Fundamental Sciences, 19395-5531 Tehran, Iran
\and
European Southern Observatory, Alonso de Cordova 3107, Vitacura, Santiago, Chile
\and
GEPI, Observatoire de Paris, PSL Research University, CNRS, Universit\'e Paris-Diderot, Sorbonne Paris Cit\'e, Place Jules Janssen, 92195 Meudon, France
\and
Instituto de Astrof\'isica de Canarias (IAC), E-38205 La Laguna, Tenerife, Spain
\and
Universidad de La Laguna, Dpto. Astrof\'isica, E-38206 La Laguna, Tenerife, Spain
\and
School of Chemistry, The University of Nottingham, University Park, Nottingham NG7 2RD, UK
\and
Royal Astronomical Society, Burlington House, Piccadilly, London W1J 0BQ, UK
\and
AAAS Science International, Clarendon House, Clarendon Road, Cambridge CB2 8FH, UK
\and
UK Astronomy Technology Centre, Royal Observatory, Blackford Hill, Edinburgh, EH9 3HJ, UK
\and
Instituut voor Sterrenkunde, KU\,Leuven, Celestijnenlaan 200D, bus 2401, Leuven, Belgium
\and
Sackler Laboratory for Astrophysics, Leiden Observatory, Leiden University, PO Box 9513, NL2300 RA Leiden, The Netherlands
\and
Astrochemistry Laboratory, NASA Goddard Space Flight Center, Code 691, 8800 Greenbelt Road, Greenbelt, MD 20771, USA
\and
Department of Physics, The Catholic University of America, Washington, DC 20064, USA
\and
Lennard-Jones Laboratories, Keele University, ST5 5BG, UK
\and 
ESTEC, ESA, Noordwijk, The Netherlands
\and
ICCM, Madrid, Spain
\and
Anton Pannekoek Institute for Astronomy, University of Amsterdam, NL-1090 GE Amsterdam, The Netherlands
\and
George Washington University, Washington DC, USA
\and
LERMA, Observatoire de Paris, PSL Research University, CNRS, Sorbonne Universit\'es, UPMC Universit\'e Paris 06, 92190 Meudon, France
\and
INAF - Osservatorio Astronomico di Cagliari, Via della Scienza 5, I-09047 Selargius, Italy
\and
NASA Ames Research Center, Space Science \& Astrobiology Division, Moffett Field, California, USA
\and
Kapteyn Institute, University of Groningen, Groningen, The Netherlands
}

\authorrunning{N.L.J.~Cox et al.}

\date{Received 3 April 2017; Accepted 25 July 2017}

\abstract{
The carriers of the diffuse interstellar bands (DIBs) are largely unidentified molecules ubiquitously present in the interstellar medium (ISM). After decades of study, two strong and possibly three weak near-infrared DIBs have recently been attributed to the C$_{60}^+$ fullerene based on observational and laboratory measurements. There is great promise for the identification of the over 400 other known DIBs, as this result could provide chemical hints towards other possible carriers. 

In an effort to systematically study the properties of the DIB carriers, we have initiated a new large-scale observational survey: the ESO Diffuse Interstellar Bands Large Exploration Survey (EDIBLES). The main objective is to build on and extend existing DIB surveys to make a major step forward in characterising the physical and chemical conditions for a statistically significant sample of interstellar lines-of-sight, with the goal to reverse-engineer key molecular properties of the DIB carriers.

EDIBLES is a filler Large Programme using the Ultraviolet and Visual Echelle Spectrograph at the Very Large Telescope at Paranal, Chile. It is designed to provide an observationally unbiased view of the presence and behaviour of the DIBs towards early-spectral-type stars whose lines-of-sight probe the diffuse-to-translucent ISM. 
Such a complete dataset will provide a deep census of the atomic and molecular content, physical conditions, chemical abundances and elemental depletion levels for each sightline.
Achieving these goals requires a homogeneous set of high-quality data in terms of resolution ($R \sim 70\,000$ -- $100\,000$), sensitivity (S/N up to 1000 per resolution element), and spectral coverage (305--1042~nm), as well as a large sample size (100$+$ sightlines). In this first paper the goals, objectives and methodology of the EDIBLES programme are described and an initial assessment of the data is provided. 
}


\maketitle

\section{Introduction}

The unknown identity of the carriers of all but two diffuse interstellar bands (DIBs) constitutes the longest standing spectroscopic enigma of modern astronomy (\citealt{2006JMoSp.238....1S}). Two features at 5797 and 5780\,\AA , which are now known to be interstellar in origin, were first noted by \citet{1922LicOB..10..141H} and studied in relation to interstellar gas and dust by \citet{1938ApJ....87....9M}. At present, over 400 of these interstellar absorption features are known (for a handful of sightlines), superimposed on an otherwise nearly smooth 
interstellar extinction curve (\citealt{1995ARA&A..33...19H}; \citealt{2000PASP..112..648G}; \citealt{2008ApJ...680.1256H}). Only recently has the attribution of a pair of near-infrared DIBs (\citealt{1994Natur.369..296F}) to C$_{60}^+$ been confirmed with laboratory gas phase experiments (\citealt{2015Natur.523..322C}; \citealt{2016NatCo...713550K}) along with the tentative astronomical detection of three more predicted bands (\citealt{2015ApJ...812L...8W,2016ApJ...831..130W}), though this needs further verification and investigation (\citealt{2017MNRAS.465.3956G,2017arXiv170401501C}). 
This is an exciting result because C$_{60}$ (\citealt{2010Sci...329.1180C}; \citealt{2010ApJ...722L..54S}), and C$_{60}^+$ (\citealt{2013A&A...550L...4B}) have also recently been detected in space through their mid-infrared emission spectra. 
This identification may be a chemical clue towards identifying further DIB carriers; so far only C$_3$ (\citealt{1995ApJ...453..450H}; \citealt{2001ApJ...553..267M}; \citealt{2014MNRAS.441.1134S}) and C$_{60}^+$ have been identified as pure polyatomic carbon species in the diffuse ISM and this leaves a large gap to be filled in our current understanding of the carbon chemical network in diffuse clouds. It is possible that the detection of C$_{60}^+$ hints at a long predicted important role of  polycyclic aromatic hydrocarbons (PAHs) in the ISM (\citealt{1985A&A...146...76V}; \citealt{1985A&A...146...81L}; \citealt{1996ApJ...458..621S}). 
Recent laboratory (\citealt{2014ApJ...797L..30Z}) and modelling (\citealt{2015A&A...577A.133B}) works support the proposal by \citet{2012PNAS..109..401B} that fullerenes may form upon photo-dissociation of large PAH precursors.

Observational surveys (e.g. \citealt{1993ApJ...407..142H}; \citealt{2011ApJ...727...33F}; \citealt{2013ApJ...774...72K}) have shown that the strength of the strongest $\sim$20 of the DIBs correlates roughly linearly with the amount of dust and gas, measured by the reddening \Ebv\ or the column density of atomic hydrogen $N$(\ion{H}{i}), respectively. This indicates a thorough mixing of the DIB carriers with interstellar matter (\citealt{2011EAS....46..349C}). 
However, a large real scatter is observed in these relations with gas and dust, and the relative strengths of several bands are known to have an environmental dependence (\citealt{2006A&A...451..973C}); some bands vary as a function of radiation strength between different lines-of-sight (\citealt{1987ApJ...312..860K}; \citealt{1997A&A...326..822C}; \citealt{2011A&A...533A.129V}; \citealt{2011ApJ...727...33F}). The scatter is also partly due to multiple cloud structures along sightlines. 
The relationship between several DIBs and, for example, C$_2$ and CN, has been investigated (\citealt{2003ApJ...584..339T,2008A&A...484..381W}), but generally the link with di-atomic species is not well understood.
This relation between DIBs and reddening has not been investigated for the remaining $>$380 bands. Whether or not there is a direct physical connection between DIB carriers and dust grains, e.g. in terms of depletion onto grains or as carrier formation sites, remains to be seen -- so far no polarisation signal has been detected for the twenty strongest DIBs (\citealt{2011A&A...532A..46C}).

Studies of selected bands in a dozen sightlines have revealed a complex substructure in the narrowest bands (\citealt{1995MNRAS.277L..41S}; \citealt{1996A&A...307L..25E}; \citealt{2008ApJ...682.1076G}) which show small variations with local temperature (\citealt{2004ApJ...611L.113C}; \citealt{2010MNRAS.402.2548K}), typical for a molecular carrier. 
Substructure has also been identified in weak DIBs, but line-of-sight variations are less well studied (\citealt{2005ApJ...629..299G}).
On the other hand, broader DIBs do not contain substructure (\citealt{2002ApJ...567..407S,2008ApJ...682.1076G}), which may be due to lifetime broadening of the absorption band (\citealt{2010A&A...511L...3L}).
Considerations of the available elemental abundances and plausible oscillator strengths lead to the conclusion that abundant large organic molecules are suitable candidates (\citealt{1985A&A...146...81L}; \citealt{2015MolPh.113.2159H}). The combination of observational studies, theoretical models, and laboratory astrophysics indicates that candidate carriers should primarily be sought among a large number of possible carbon-based organic molecules [see \citet{2006JMoSp.238....1S} for a review, and \citet{2014IAUS..297.....C} for an overview of recent progress].

Ongoing and future large spectroscopic surveys offer the possibility to study (mostly the strongest) DIBs in large areas of the sky. For example, \citet{2014ApJ...795...31L} and \citet{2015MNRAS.451..332B} constructed DIB  strength maps from SDSS spectra, \citet{2014Sci...345..791K} produced pseudo-3D maps for the 8621~\AA\ DIB using the RAVE survey, \citet{2015ApJ...798...35Z} and \citet{2016ApJS..225...19E} used the APOGEE near-infrared survey to study the distribution of the 15267~\AA\ near-infrared DIB. The spatial distribution and properties of DIBs can also be studied in smaller fields-of-view (\citealt{2009MNRAS.399..195V}; \citealt{2012A&A...544A.136R}; \citealt{2015A&A...573A..35P}) or closer regions, such as the Local Bubble
(\citealt{2015ApJ...800...64F,2016A&A...585A..12B}).

In the last decade it has also been firmly established that many band carriers are universal; DIBs have been detected and surveyed in the Magellanic Clouds (\citealt{2006A&A...447..991C,2007A&A...470..941C,2006ApJS..165..138W,2013A&A...550A.108V,2015MNRAS.454.4013B}), in M31 and M33 (\citealt{2008A&A...492L...5C,2011ApJ...726...39C}), and in individual sightlines in more distant galaxies (\citealt{2004ApJ...614..658J}; \citealt{2005A&A...429..559S}; \citealt{2008AJ....136..994L}; \citealt{2008A&A...485L...9C,2014A&A...565A..61C}; \citealt{2015A&A...576L...3M}). DIB carriers therefore constitute an important reservoir of (organic) material throughout the Universe.

Identifying the DIB carriers and understanding their properties must come from high-quality data in the nearby Galactic interstellar medium (ISM). Identification of the carrier species will directly impact our understanding of interstellar chemistry, and can help reconstruct 3D line-of-sight properties if related to specific environments.
It is clear that the ultimate confirmation must come from a direct comparison between astronomical, theoretical, and laboratory spectra over a broad wavelength range. A commonly applied and straightforward approach is to acquire laboratory spectra of possible candidate carriers taken under astrophysical relevant conditions until an unambiguous match with the astronomical data is found. With the notable exception of the above mentioned work on C$_{60}^+$ previous studies have thus far failed, such as attempts to link PAH cations (\citealt{1999A&A...343L..49B,2011ApJ...728..154S,1999ApJ...526..265S}; \citealt{1999CPL...303..165R}), neutral PAHs (\citealt{2011ApJ...728..154S,2011A&A...530A..26G}), carbon chains (\citealt{2000ApJ...531..312M}; \citealt{2004ApJ...602..286M}), or H$_2$ (\citealt{1999h2sp.confE..68S}; \citealt{2014IAUS..297..375U}) to the DIBs. The search for a laboratory match can be optimised if the most likely candidates can be pre-selected out of the vast collection of possible species, and if the relevant conditions can be accurately constrained. Hence it is necessary to unravel the physical and chemical properties of the DIB carriers through analysis and modelling of observations.
This includes deriving environmental conditions that affect their strength/profile shapes, as well as understanding the molecular physics and spectroscopy of candidate carriers.

This paper presents the observational overview of the ESO Diffuse Interstellar Bands Large Exploration Survey (EDIBLES) and how we intend to use the obtained spectra in our long-term goal of reverse-engineering the molecular characteristics of DIB carriers. In Sect.~\ref{sec:goals-objectives} we describe the scientific goals and immediate objectives of EDIBLES. Sect.~\ref{sec:methodology} describes the methodology and survey design. The survey target selection is discussed in Sect.~\ref{sec:targets} and the data processing steps are described in Sect.~\ref{sec:processing}. Sect.~\ref{subsec:telluric} discusses several confounding factors such as telluric and stellar spectral lines.
In Sect.~\ref{sec:preview} we present a preview of the EDIBLES data and illustrate their scope and quality. A brief summary is given in Sect.~\ref{sec:summary}.

\begin{table}[t!]
\caption{UVES instrument setups used.
The two spectrograph arms are used to collect data for a pair of wavelength regions simultaneously. A second setup allows the gaps to be covered with another pair of settings. For each setting we give the slit width, nominal resolving power, and nominal wavelengths covered.
Red-L and Red-U refer to the spectra recorded with the Red Lower EEV CCD and the
Red Upper MIT CCD (c.f. Sect.~\ref{sec:processing}).}
\label{tb:setup}
\centering
\begin{tabular}{llcrll}\hline\hline
Setting  	 & Arm		& Slit width	& Resolving 		& Spectral range 	\\ 
(nm)	 	 &			& (\arcsec)	& power			& (nm)			\\\hline
346		 & blue		& 0.4			& 71\,000			& 304.2 -- 387.2	\\
564		 & red-L		& 0.3			& 107\,000		& 461.6 -- 560.8 	\\
		 & red-U		&			&				& 566.9	-- 665.3	\\
\hline	
437		 & blue		& 0.4			& 71\,000			& 375.2 -- 498.8	\\
860		 & red-L		& 0.3			& 107\,000		& 670.4 -- 853.9	\\ 
		 & red-U		&			&				& 866.0 -- 1042.0	\\

\hline	
\end{tabular}
\end{table}

\section{Scientific goals \& immediate objectives}\label{sec:goals-objectives}

The primary science goal of EDIBLES is to reverse-engineer molecular characteristics of DIB carriers, through studying the behaviour of DIBs in relation to the physical and chemical parameters of their environment. This approach differs from earlier work in which attempts to identify DIBs were based mainly on direct comparisons of astronomical and laboratory or theoretical spectra. A large systematic high-fidelity survey of the diffuse-to-translucent ISM is necessary to realise this approach.

The aim is to assemble a sample of interstellar spectra with sufficiently high spectral resolution and signal-to-noise ratio to allow detailed analysis of numerous DIBs and known atomic and molecular absorption lines in the same lines-of-sight. At the same time, our sample is designed to sample a wide range of interstellar conditions, in terms of reddening, molecular content and radiation field, within a practical observing time.

With EDIBLES we plan to compile the global properties of a large ensemble of both weak and strong DIBs, and variations therein, as a function of depletion (patterns) and local physical conditions. The new dataset should allow us to: 
(a) determine the relation between weak and strong DIBs by identifying correlations and sequences;
(b) identify (sets of) DIBs that correlate with different physical conditions in the ISM, and assess whether the DIBs can be used to determine those conditions as a remote diagnostic tool;
(c) study the physico-chemical parameters that influence the DIB properties, by using state-of-the-art chemical modelling, combined with extensive auxiliary line-of-sight data (e.g. on dust); and
(d) constrain the chemical composition of the DIB carriers by studying their relation to interstellar elemental abundances (depletion levels) and dust grain properties and composition derived from, for example, 
the UV-visual extinction (including the conspicuous 2175~\AA\ UV bump) and optical polarisation curves.

A number of studies have attempted to investigate links between the physical and chemical conditions of the ISM and the properties of the DIBs. However, most studies focus only on a few strong bands in a moderate-to-large number ($\approx$100) of sightlines (\citealt{2011ApJ...727...33F}; \citealt{2011A&A...533A.129V}; \citealt{2015ApJ...798...35Z}), or on many DIBs in just a few sightlines (\citealt{1997A&A...326..822C}; \citealt{2000A&AS..142..225T}; \citealt{2008ApJ...680.1256H, 2009ApJ...705...32H}).  Hence, more recent progress in the field has been limited to the study of only a handful of the strongest DIBs due to high demands on the signal-to noise ratio (S/N), spectral resolution, the removal of stellar and telluric lines, and the lack of large, uniform data sets. 
EDIBLES is designed to fill this gap, making just such a large, uniform data set available and thus enabling a large and systematic study
of the physical and chemical parameters that are expected to directly influence the formation efficiency and spectroscopic response of DIB carriers. 

\section{Methodology \& survey design}\label{sec:methodology}

EDIBLES provides the community with optical ($\sim$305--1042~nm) spectra at high spectral resolution 
(R $\sim $ 70\,000 in the blue arm and 100\,000 in the red arm) and high signal-to-noise (S/N; median value $\sim$ 500--1000), 
for a statistically significant sample of interstellar sightlines. Many of the $>$100 sightlines included in the survey already have auxiliary available ultraviolet, 
infrared and/or polarisation data on the dust and gas components.

Studies of DIBs typically report data such as equivalent width, central depth, profile shape and substructure identification. These cannot easily be compared between surveys due to differences in the instrumentation, data quality and analysis procedures -- e.g. continuum normalisation and measurement of spectroscopic lines. Archival material comprises a heterogeneous sample of spectra with varying S/N, resolving power, and spectral coverage. To achieve the goals and objectives described above requires a large and homogeneous survey of UV/visible spectroscopic tracers across a broad spectral range, covering a broad variety of interstellar environments. From this self-consistent set of observations we can extract:

\begin{enumerate}

\item Accurate column density measurements (or upper limits) for the most important atomic and molecular species, across a wide spectral range. These can be used to assess the velocity structure of the line-of-sight (in particular to determine radial velocity differences between species; e.g. \citealt{2007MNRAS.378..893B}), derive depletion levels of metals, infer and compute physical conditions (within the limitations imposed by the current knowledge on interstellar processes), using diffuse cloud PDR models (\citealt{2006ApJS..164..506L}) or turbulent energy dissipation models (c.f. \citealt{Flower15}; \citealt{Godard14}; \citealt{EmericThese}).

For the photo-chemistry and derivation of particle density, radiation fields, turbulent energy dissipation, the key transitions are those of CN $\lambda\lambda$3874, 7906, CH $\lambda\lambda$3879, 4300, and CH$^+$ $\lambda\lambda$3958, 4232~\AA, together with H$_2$ (from archival UV spectroscopy). 
For example, rotational temperatures can be derived from bands of C$_2$ ($\lambda8756$) and C$_3$ ($\lambda$4053), and cosmic ray ionisation rates can be derived from OH$^+$ abundances ($\lambda\lambda$3300--3600~\AA).
CH measurements can be used to estimate the H$_2$ column density (\citealt{1984A&A...130...62D,2004A&A...414..949W}).

\item Accurate measurements of DIB profiles (asymmetries, wings, substructure) and variations therein. Substructure can be related to molecular properties / sizes of carrier species (\citealt{1996MNRAS.283L.105K,1996A&A...307L..25E,2015MolPh.113.2159H}) with variations due to changes in the rotational temperature (\citealt{2004ApJ...611L.113C}; \citealt{2010MNRAS.408.1590K}) or the presence of hot bands (\citealt{2015MNRAS.453.3912M}).

\item Updated measurements of peak positions of weak (per unit reddening) diffuse bands along single cloud sightlines. 

\item Measurements and cross-correlation of over 50 weak and strong 
bands along the most reddened  sightlines (\Ebv~$>$~0.4~mag). Correlations between strong and weak bands might reveal additional information on groups (or families) of DIBs, but it should be noted that a strong correlation between DIBs is not a necessarily a guarantee that they have a common carrier (\citealt{2010ApJ...708.1628M,2016MNRAS.460.2706K}).

\item Stacking analyses to search for molecules and/or DIBs which are too weak to be seen in individual spectra.

\item Firm detection limits or abundance constraints on specific molecular carriers for which laboratory spectra are obtained. 

\item Variations in interstellar species due to the small-scale structure of the diffuse ISM (\citealt{2013ApJ...764L..10C,2013MNRAS.429..939S}).

\end{enumerate}

\begin{figure*}[th!]
\centering
\includegraphics[width=0.9\textwidth,viewport=10 0 640 380, clip]{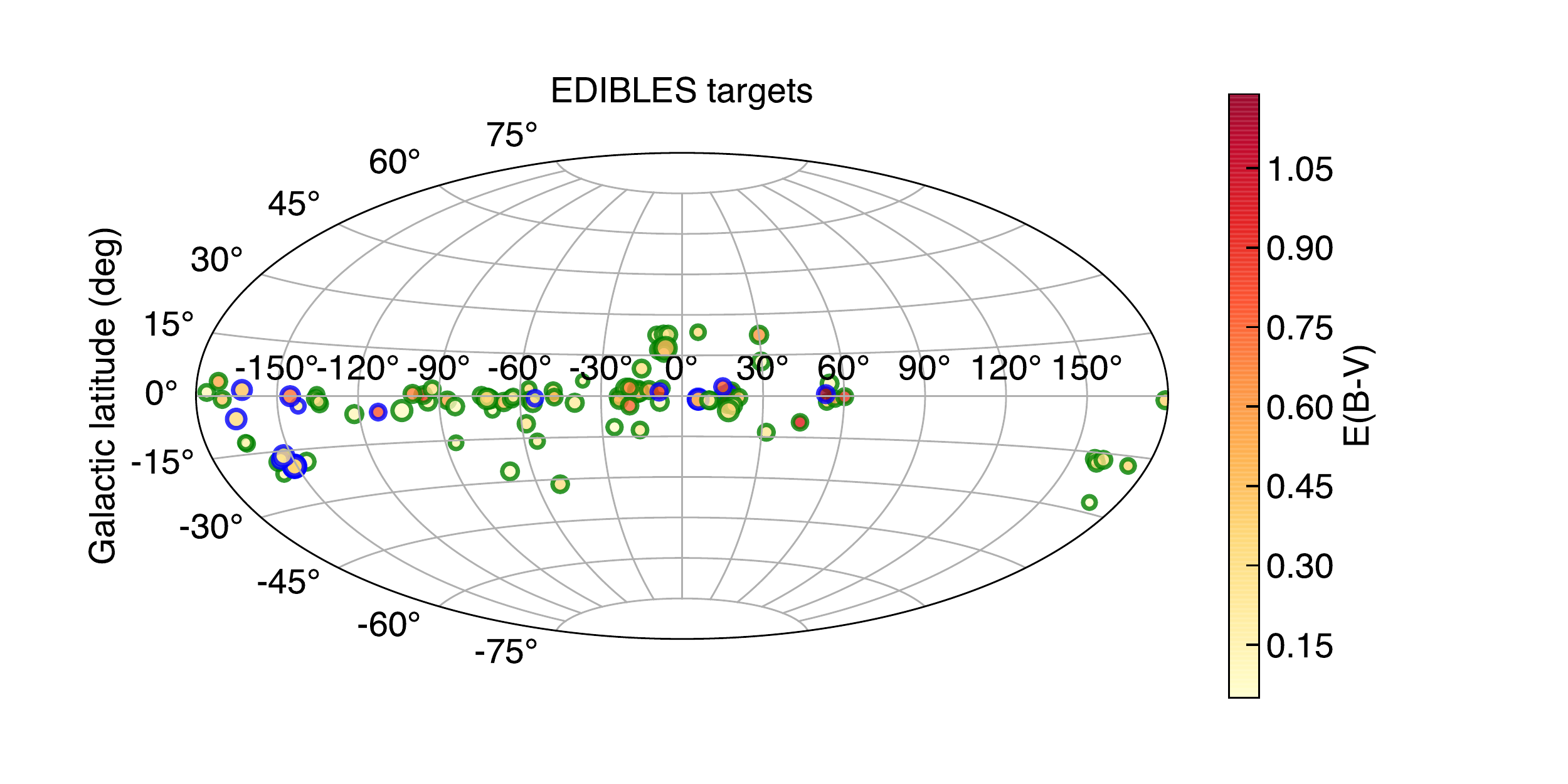}
\caption{Galactic distribution of EDIBLES targets.
The symbol size reflects the value of $R_V$, while the interior colour represents the line-of-sight reddening, \Ebv.
Symbols with green edges represent the observed targets, while blue edges correspond to the targets to be observed by the end of the programme.}
\label{fig:aitoff}
\end{figure*}

To achieve our objectives efficiently we use the Ultra-violet Visual Echelle Spectrograph (UVES; \citealt{2000SPIE.4008..534D}; \citealt{2009Msngr.138....8S}) mounted on the 8-metre second Unit Telescope (UT2) of the ESO (European Southern Observatory) Very Large Telescope at the Paranal Observatory. The relative brightness of nearby early-type stars allows the observation strategy to take advantage of poor observing conditions and twilight hours that would otherwise be under-utilised. The programme is running as a Large `Filler' Programme (ESO ID 194.C-0833, PI. N.L.J. Cox), which has been allocated 280 hours of observing time. About 8\,500 science exposures with a total exposure time of 229 hours (with blue and red arm exposures taken simultaneously) have been collected between September 2014 and May 2017. The program is expected to be completed by late 2017.

UVES has two arms, red and blue, which can be used simultaneously by inserting a dichroic mirror \citep{2000SPIE.4008..534D}. To obtain coverage of the entire spectral range accessible with UVES, we use two instrumental settings per target: setting \#1: 346+564 and setting \#2: 437+860, where the pairs of numbers refer to the central wavelengths in nanometres of the two arms.
Together this provides near-continuous wavelength coverage from $\sim$305 to 1042~nm. 
The blue and red arm slit widths are 0.4\arcsec\ and 0.3\arcsec, respectively, yielding 
nominal resolving powers $R=\lambda/\Delta\lambda$ of $\sim$71\,000 and $\sim$107\,000. Table~\ref{tb:setup} presents a summary of the instrumental setups. 
The EDIBLES data presented here were collected over a period of two years, and it is therefore important to realize that the UVES resolution is not fully stable with time\footnote{\url{http://www.eso.org/observing/dfo/quality/UVES/reports/ HEALTH/trend\_report\_ECH\_RESOLUTION\_DHC\_HC.html}}, but the values listed in Table~\ref{tb:setup} were typically realised in the actual spectra.

The ``filler''-type observation strategy means that observations are often executed in non-optimal (and unpredictable) conditions e.g. in terms of seeing, cloud coverage, sky emission/air glow, lunar phase, and water vapour content. This needs to be taken into account in the implementation of the observations. Despite these limitations, S/N ratios of 200--300 per exposure can be reached in short exposure times for the bright ($2 < V < 6$~mag) targets, and up to 20 exposures (to avoid saturation of individual frames) are obtained for each to build up higher S/N. Observations are divided into observing blocks (OBs) for a specific instrument setting and target. OB execution times range from $\sim$20 minutes for the bright ($V < 6$ mag) stars up to $\sim$45 minutes for the fainter ($6 < V < 9$) stars.

Additional flat-field calibration exposures were taken during the day-time (when possible, subject to operational constraints) to reduce residual fringing that would persist with the standard UVES calibration plan and to increase the overall S/N ratio. Details of the flat-field corrections are given in Sect.~\ref{sec:processing}.

\begin{figure*}[ht!] \centering
\includegraphics[width=0.4\textwidth, viewport=30 10 470 350, clip]{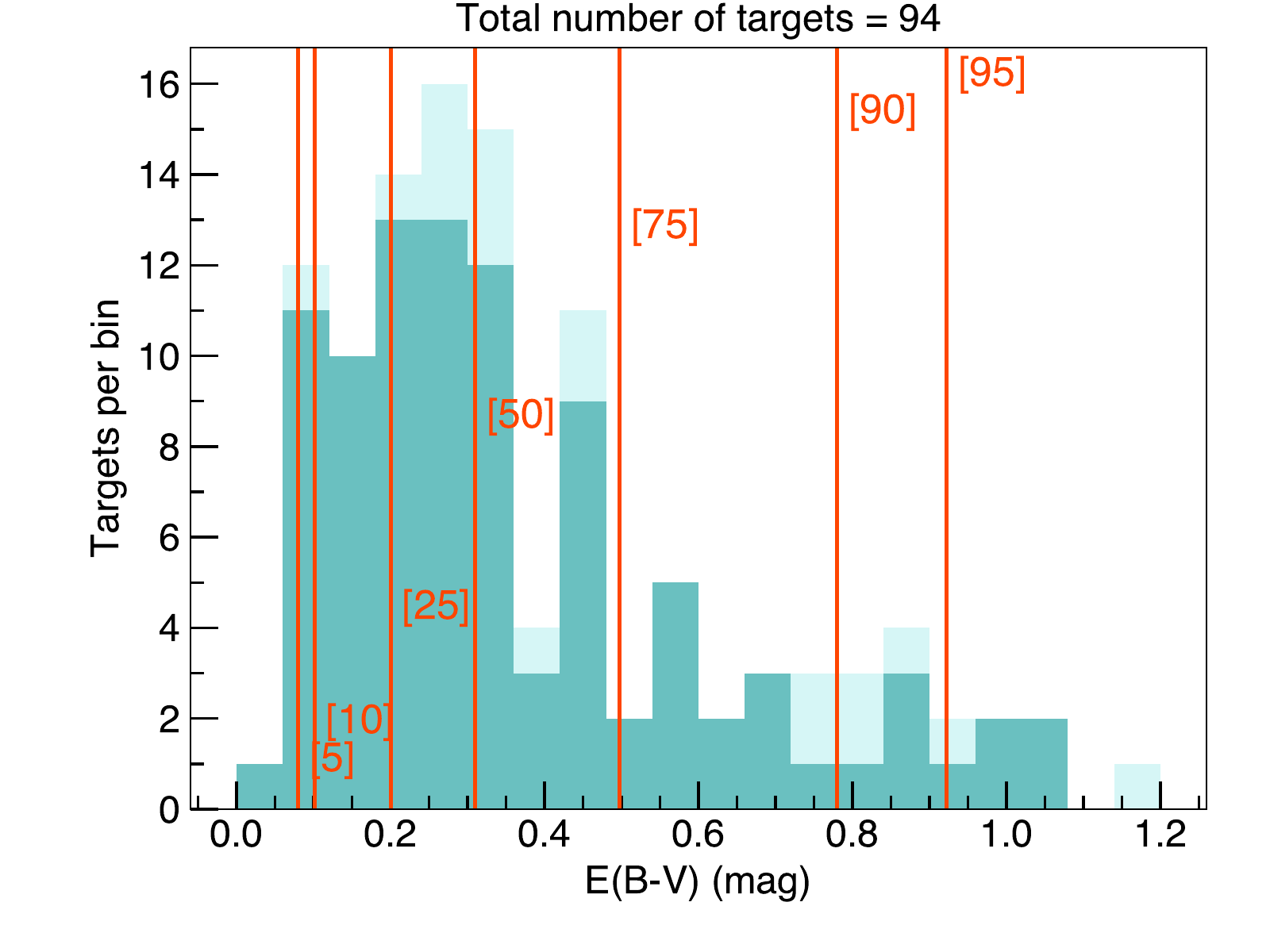}
\includegraphics[width=0.4\textwidth, viewport=30 10 470 350, clip]{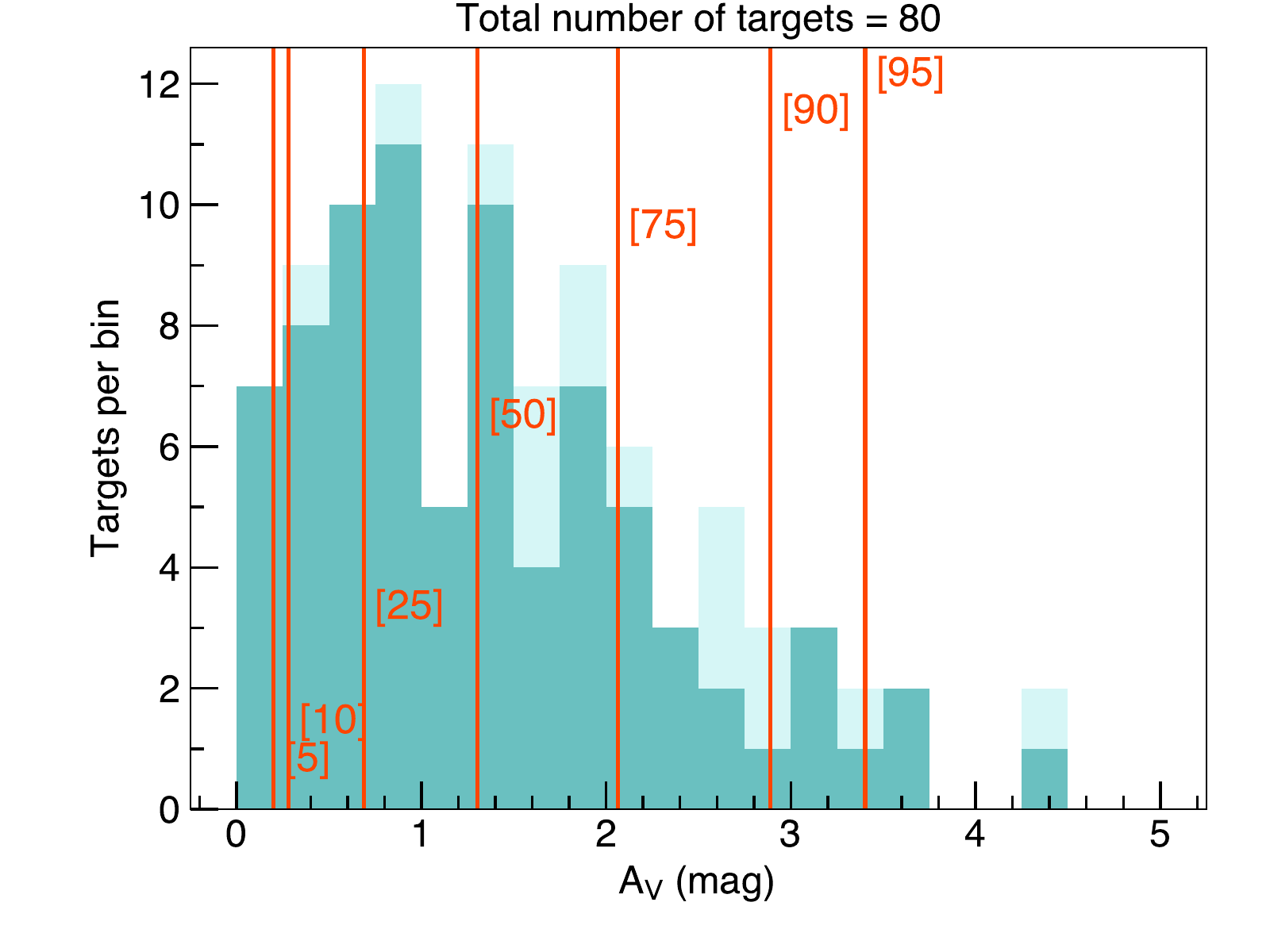}\\
\includegraphics[width=0.4\textwidth, viewport=30 10 470 350, clip]{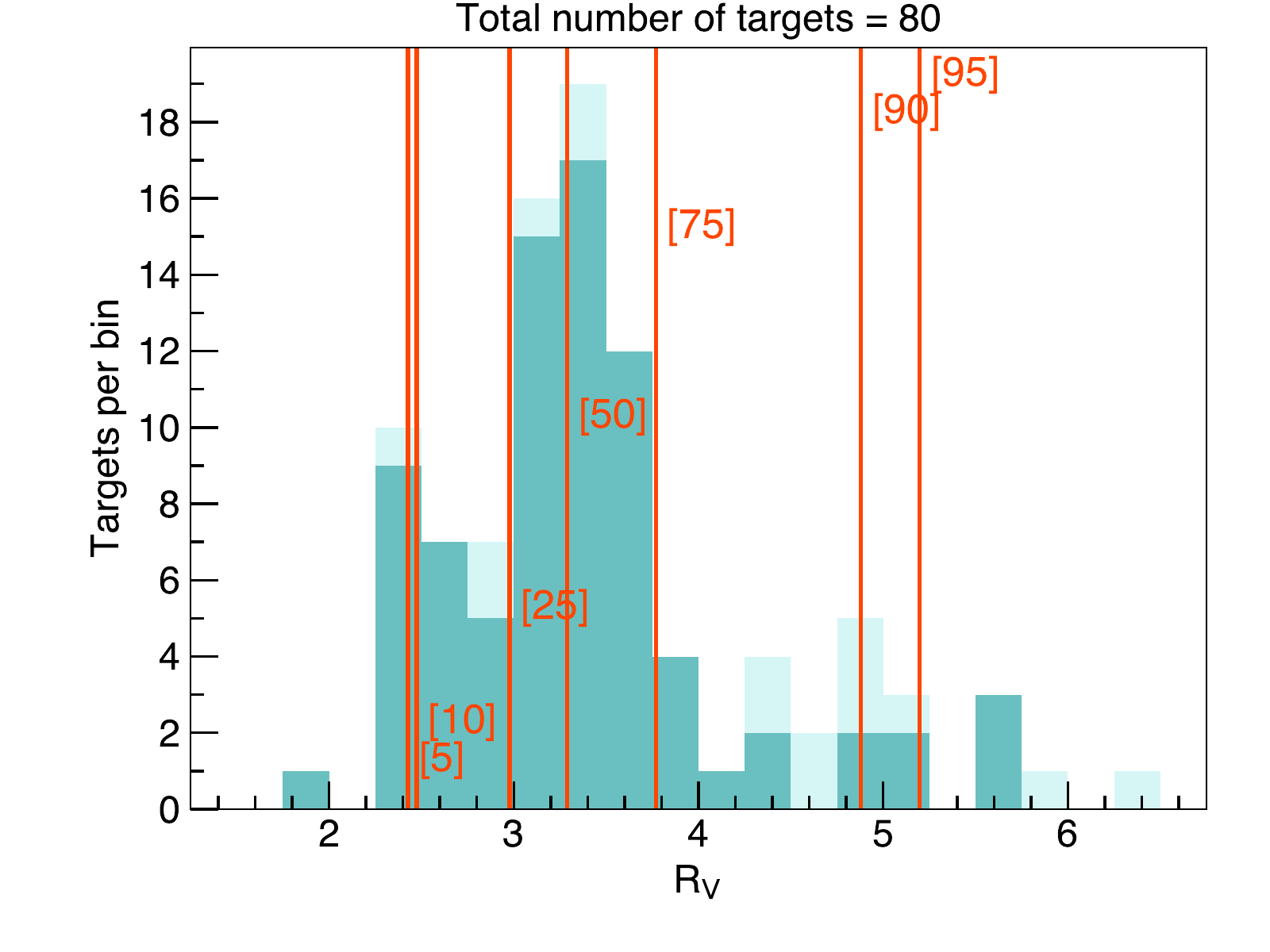}
\includegraphics[width=0.4\textwidth, viewport=30 10 470 350, clip]{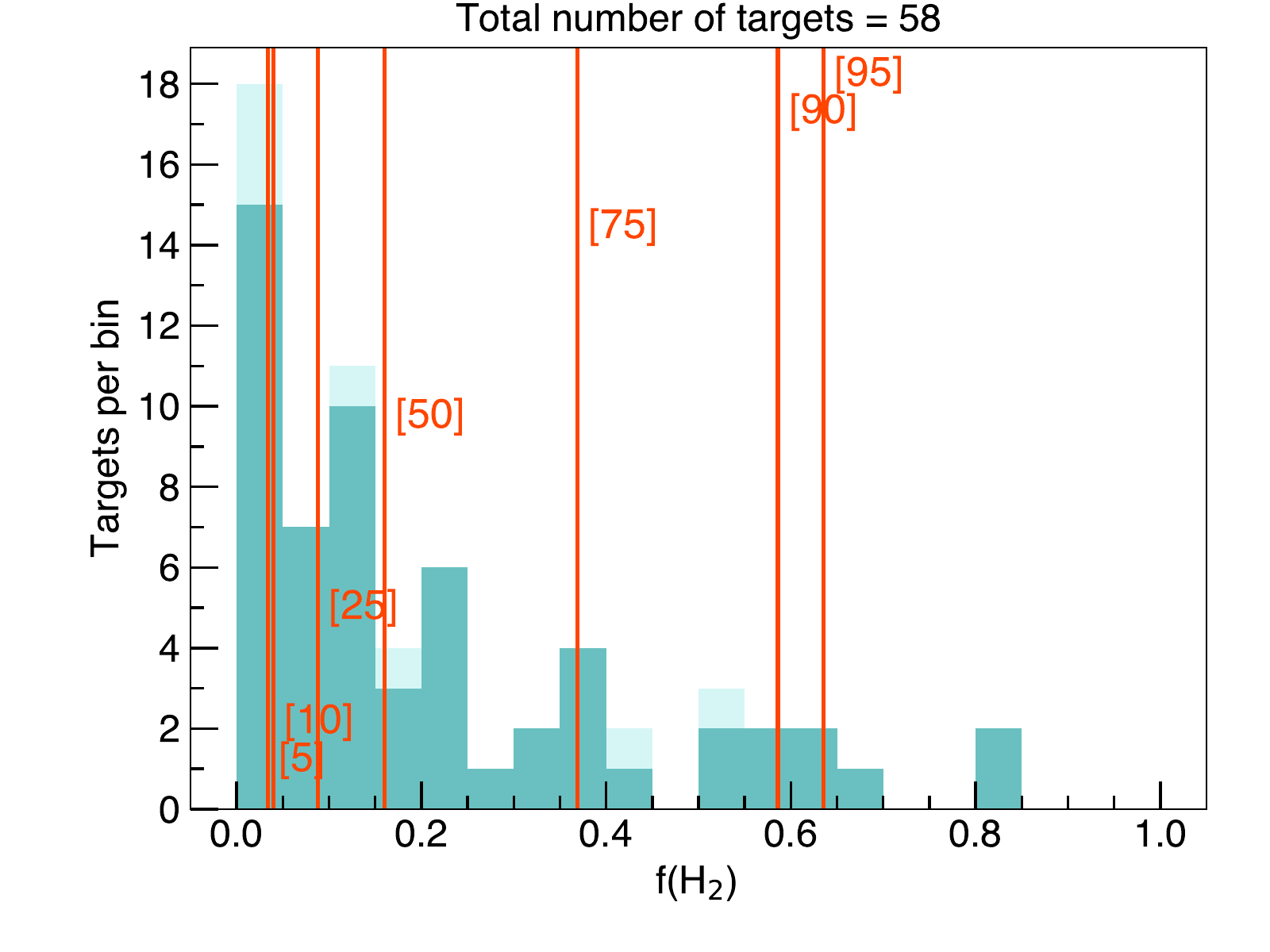}
\caption{Number of selected targets as function of reddening \Ebv, extinction $A_V$, the ratio of total-to-selective extinction $R_V$ ($=A_V$/\Ebv), and the fraction of molecular hydrogen \fHtwo for the target sample.
The number of observed targets with reported values for each quantity are given at the top of each panel. The dark blue and light blue distributions correspond to the samples of observed and observed $+$ foreseen targets.
The vertical red lines indicate the value of the 5, 10, 25, 50, 75, 90, and 95 percentiles of each sample.
The labels are located  such that they trace the cumulative target distribution.
}
\label{fig:AV}\label{fig:RV}\label{fig:fH2}
\end{figure*}

\section{Target survey sample selection and characteristics}\label{sec:targets}

We constructed a statistically representative survey sample that probes a wide range of interstellar environment parameters including reddening \Ebv, visual extinction $A_V$, total-to-selective extinction ratio $R_V$, and molecular hydrogen fraction \fHtwo. This is essential to (a) trace depletion patterns from diffuse $\rightarrow$ translucent clouds, (b) study the effect of shock- and photo-processing, (c) probe the behaviour of DIBs with respect to grain properties, and (d) identify unusual DIB environments.

During target selection the following factors were taken into account:

\begin{itemize}

\item Given that \fHtwo depends non-linearly on $A_V$, due to the transition from atomic to molecular hydrogen driven by H$_2$ self-shielding, we require numerous sightlines probing $A_V \sim 1 - 3$~mag and below, in small increments $\Delta$$A_V$. 

\item The dust grain properties and attenuation of UV photons (important for photo-chemistry) are constrained by the extinction curve, i.e. the $A_V$ and $R_V$ parametrisation (\citealt{2004ApJ...616..912V}; \citealt{2007ApJ...663..320F}) or from fitting with a well-defined dust-PAH extinction model (\citealt{2013ApJS..207....7M}). 

\item Preference is given to sightlines with auxiliary atomic/molecular data, such as \ion{H}{i} and H$_2$ measurements (\citealt{2009ApJ...700.1299J}; \citealt{2012ApJS..199....8G}), optical polarisation data (\citealt{1992ApJ...386..562W}; \citealt{2008AcA....58..433W}), or Mg/Fe abundances (\citealt{2012A&A...541A..52V}).

\end{itemize}

\begin{figure}[ht!]
\centering
\includegraphics[width=0.95\columnwidth]{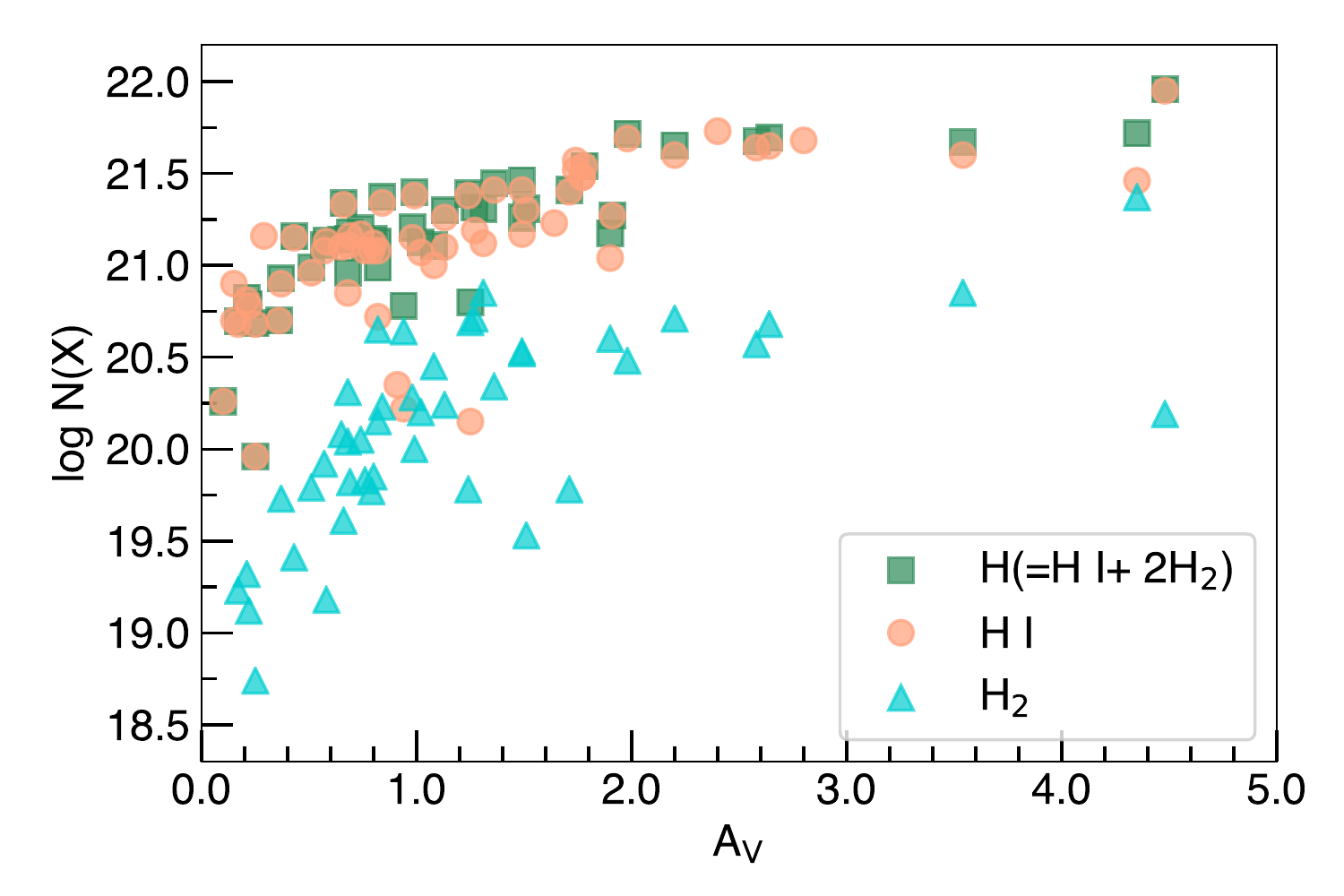}
\caption{Relation between visual extinction, $A_V$, and neutral hydrogen column density $N$(\ion{H}{i}), molecular hydrogen column density $N$(H$_2$), and total hydrogen column density $N$(H$_\mathrm{tot}$), computed as $N$(\ion{H}{i})+2$N$(H$_2$). Note that some EDIBLES lines-of-sight are not included since no direct \ion{H}{i} or H$_2$ measurements are available.}
\label{fig:hydrogen}
\end{figure}

Where two targets with similar interstellar conditions are available, we preferentially selected targets which are brighter and/or of earlier spectral type. 

The target list is given in the Appendix, Table~\ref{tb:targetlist1} (their Galactic distribution is shown in Fig.~\ref{fig:aitoff}). Columns (1) to (3) provide basic information on the target id (HD number) and coordinates (RA/Dec). Columns (4) and (5) list the spectral type and corresponding literature reference.
The interstellar line-of-sight dust extinction properties, \Ebv, $R_V$, $A_V$, are given in columns (6) to (9).
Columns (10) and (11) list the atomic and molecular hydrogen abundances with the molecular fraction \fHtwo listed in column (12).

The total number of selected targets amounts to 114 (of which 96 have been observed at least once as of May 2017) and comprises mostly bright O and B stars ($V \approx$ 2--7 mag with a small fraction $7 < V < 9$~mag).
The sample probes a wide range of interstellar dust extinction properties 
(\Ebv $\sim$ 0--2~mag; $R_V \sim$ 2--6; $A_V \sim$ 0.1--4.5~mag) 
and molecular content (\fHtwo $\sim$ 0.0--0.8).
The histograms in Fig.~\ref{fig:AV} illustrate the range of parameters 
included in the survey sample.
In Fig.~\ref{fig:hydrogen} we show comparisons between visual extinction, $A_V$, and 
the measured neutral hydrogen column density $N$(\ion{H}{i}), molecular hydrogen column density $N$(H$_2$), 
and total hydrogen column density $N$(H$_\mathrm{tot}$) (=$N$(\ion{H}{i})+2$N$(H$_2$)).
As noted above H$_2$ can be estimated using the CH transitions  (\citealt{1984A&A...130...62D,2004A&A...414..949W}).

\section{Data processing}\label{sec:processing}

Acquiring high-S/N spectra with UVES is challenging in the context of EDIBLES for a number of reasons. For example, small errors in the wavelength calibration at the edges of individual adjacent orders can cause the appearance of ripples in the continuum in high-S/N exposures. Moreover, the unpredictable seeing and other observing conditions inherent in a filler programme mean that individual exposure times cannot be optimised for the actual sky conditions.

The overall quality of all 8\,500 science and $\sim$~1600 flat exposures was checked visually. 
This visual inspection led us to discard about 150 exposures that appeared corrupted. 
This could, for example, be due to sudden changes in weather conditions or premature termination of exposures. 
In addition, we carefully inspected for the presence of incorrect thorium-argon lamp, format-check, 
and flat field exposures that could result in faulty order tracing or wavelength calibration solutions (see also below).

Within a sequence of observations (20--40 minutes including overheads) the change in barycentric velocity correction is small ($<0.1$~\kms\ per hour) so the spectra can be averaged without compromising the velocity precision. 
However, exposures taken on different nights were not averaged. 
This is to preserve multi-epoch information -- specifically for spectroscopic binaries and the search for time-variable interstellar absorption -- and to avoid addition of misaligned interstellar features due to variations in the barycentric velocity of the frame of the observer.

The data reduction was performed by two semi-independent teams, one using version 5.7.0 of the UVES pipeline (\citealt{2000Msngr.101...31B}), 
\texttt{esorex} (version 3.12; \citealt{2015ascl.soft04003E}) integrated in a python pipeline (hereafter Reduction 'A'), 
and the other using the 4.4.8 version of the UVES pipeline (Reduction 'B').

Each set of around 20 science frames was processed with the same set of calibration frames. 
For the format check, order definition and wavelength calibrations these were the nearest in time, 
with the master bias and master flats using frames taken typically over several days or weeks.  
In general, both data reductions used similar parameters for the different pipeline recipes, but with a few differences.
Reduction `A' adopts optimal merging, applies the blaze correction and uses about 100--130 flats for each arm, 
while reduction `B' uses optimal merging and takes the nearest 140 flat field frames.

In the following we provide a detailed description of the different data processing steps and discuss their impact on the quality of the reduced spectra.  
The differences between reductions `A' and `B' are discussed in relation to the blaze correction and order merging step.

\begin{figure}[t!]
\centering
\includegraphics[width=0.9\columnwidth]{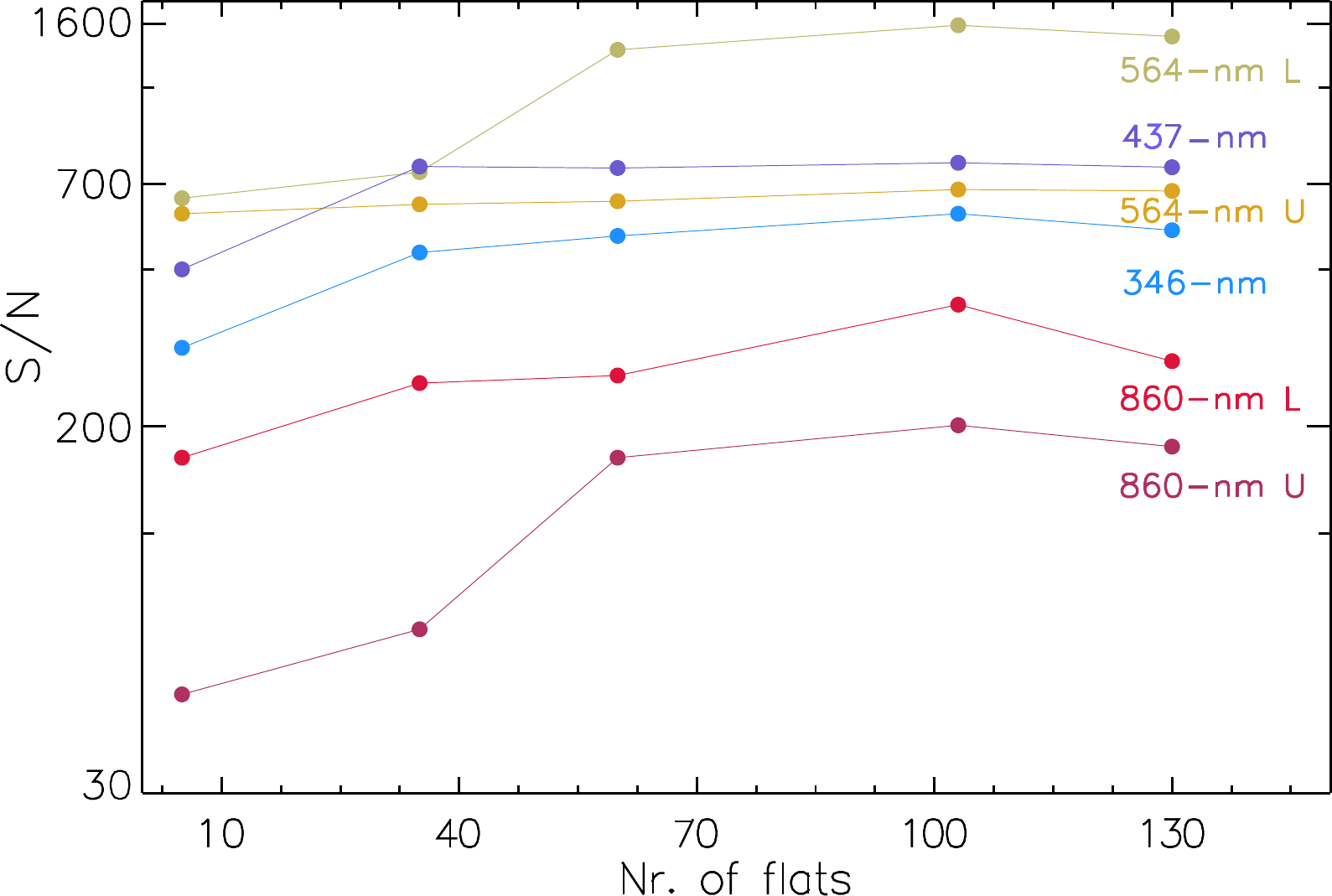}
\caption{Comparison of the S/N of HD 23180 with changing number of flat frames.
The S/N ratios plotted for each instrument setting are average values of S/N measured in five different continuum regions in the respective setting.}
\label{fig:SNR}
\end{figure}

\paragraph{Bias}
To subtract the CCD bias level in science frames, we created a master bias frame by median-stacking a set of 50 (reduction `A') or 25 (reduction `B') bias exposures, 
using the \texttt{uves\_cal\_mbias} recipe. To handle the bad columns in the REDL CCD, by default the recipe interpolates across bad pixels but does not apply this interpolation in science frames. This inconsistency creates boxy emission-like artifacts in the reduced spectra. To avoid this from occurring, high-quality bias frames are carefully pre-selected and the data are processed without interpolating over bad pixels. 

\begin{figure}[t!]
\centering
\includegraphics[width=0.9\columnwidth]{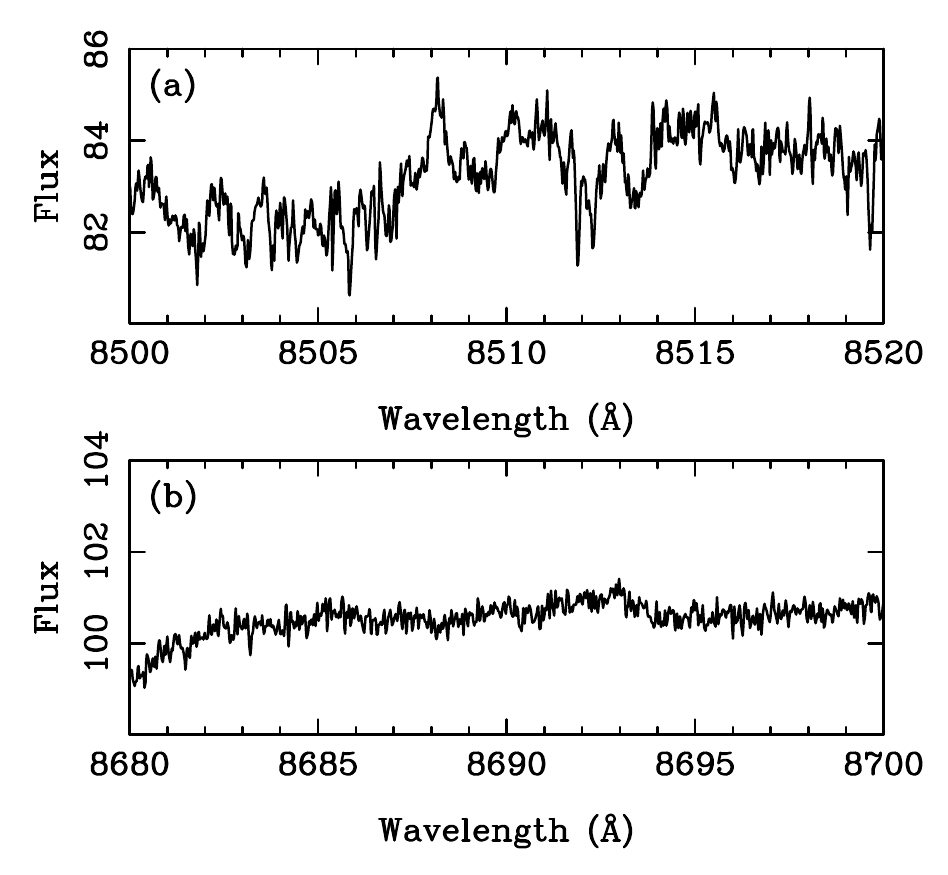}
\caption{(a) Part of the extracted spectrum of HD 23180 taken using the Red Lower EEV CCD (Red-L). 
(b) Ditto but for Red Upper MIT CCD (Red-U). The latter is a thick chip so fringing is much reduced compared with the EEV detector. The vertical scale is the same in both cases. }
\label{fig:Plot_HD23180_860nm_Amin}
\end{figure}

\paragraph{Order definition}

In order to find the physical position of echelle orders in the X and Y directions of spectral frames for a given instrument setting, \texttt{esorex} uses a physical model based on the instrument configuration, ambient pressure, the humidity, slit width, central wavelength, camera temperature and CCD rotation angle. The physical model then predicts the X and Y pixel position corresponding to the nominal orders and stores the calculations into the guess line and order tables. These tables are then normally used as the initial values for identifying the specific positions of the orders.
	
To accurately detect the order positions, \texttt{esorex} defines a search box on the detected lines in the arc lamp frames and tries to match the predicted position of the physical model lines. The following procedure robustly detects the order positions: 
\begin{enumerate}
\item measure the raw X and Y pixel positions of the thorium-argon lines on an arc frame exposure by defining a $\mathbf{60 \times 60}$ pix$^2$ square search box,
\item compute the difference of predicted and detected order positions,
\item decrease the size of the search box to $\mathbf{40 \times 40}$ pix$^2$, 
iterate the X and Y shifts to search for residuals less than $\mathbf{\pm 1}$ pixel and to reduce the root-mean-square (RMS) values,
\item fit a 2D Gaussian function to XY pixel positions within a `fit box' centred at the predicted line positions,
\item For reduction `B' additional iterations of the \texttt{format check} are done using different values of the CCD rotation offset, selecting the one with the maximum number of lines found,
\item and, finally, perform a 2D \textbf{second-order} polynomial fit in XY to the fitted line positions.
\end{enumerate}
The highlighted values in the above steps are our tuned parameters in the \texttt{uves\_cal\_predict} recipe.

\begin{figure}[t!]
\centering
\includegraphics[width=0.9\columnwidth]{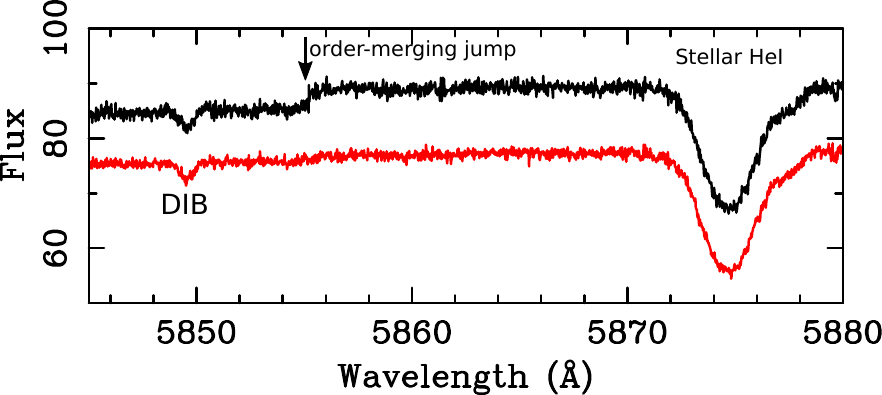}
\caption{Close-up view of a region including two overlapping orders in the 564-nm setting for HD\,23180 for both reduction `A' (bottom red trace) and reduction `B' (top black trace). The small jump in the continuum at approximately 5855~\AA\ seen in reduction B (top black trace) is due to imperfect merging of two echelle orders. 
The apparent difference in S/N is due to alternative choices of wavelength sampling.}
\label{PlotHD23180_564}
\end{figure}

\begin{table}
	\centering
	\caption{Wavelength calibration uncertainties for the thorium-argon frames corresponding each of the observed UVES settings for HD\,23180.}
	\label{tb:error}
	\begin{tabular}{lcc}\hline\hline
		 Arm & uncertainty (\AA) & uncertainty (m s$^{-1}$) \\ \hline
		346-nm    & $6.4 \times 10^{-4}$ & 52   \\
		437-nm    & $7.8 \times 10^{-4}$ & 54   \\
		564-nm L  & $4.5 \times 10^{-4}$ & 26   \\
		564-nm U  & $7.0 \times 10^{-4}$ & 34   \\
		860-nm L  & $8.3 \times 10^{-4}$ & 33   \\
		860-nm U  & $1.1 \times 10^{-3}$ & 36   \\ \hline
		
	\end{tabular}\par
\end{table}

\paragraph{Flat fielding}
The flat-fielding is applied, to the 2D frame, before the order extraction (see below). This is known as the pixel flat-fielding method.
The standard UVES pipeline uses only five flat field frames, which limits the maximum S/N.
To demonstrate the necessity for accurate flat-fielding to reach the required S/N ratio, we reduced 11 science frames of HD\,23180 using a different number of flat field frames to build the master flat frame. As shown in Fig.~\ref{fig:SNR}, increasing the number of flat fields helps to increase the final S/N. 
More than 100 flat field frames do not cause the S/N to increase further, indicating that flat fielding is no longer the limiting factor on the S/N. 
In the case of the 860-nm setting the S/N drops between 100 and 130 applied flats due to the presence of several bad-quality flat frame exposures. For the final reduction these were rejected and we selected as many as possible, usually about 100, good quality flat frames.
By using the additional flat fields, we were able to improve the final S/N by factors of two to five, depending on the wavelength.

Note that the intensity (photon counts) of normal flats are very low in the UV ($<$320~nm). Therefore, to improve the quality of the first-orders of the 346-nm arm spectra, 
a final master flat is constructed by combining a set of normal flats with flats obtained by exposing with a deuterium calibration lamp.
The merging was done at orders 145 and 146 around 321~nm with the master deuterium-lamp flat being used bluewards of this and the normal flat redwards. 
This is to avoid as much as possible spurious absorption-line features in the final science spectrum caused by emission features in the deuterium lamp redwards of 321~nm.

\begin{table}[t!]
\caption{Median S/N per ``resolution-element'' (0.04~\AA\ spectral bin) for each UVES setting/arm configuration for all available EDIBLES spectra.} 
\label{tb:snr}
\centering
\begin{tabular}{lllll}\hline\hline
Setting	& Arm	& $\lambda$-range (nm)	& Median S/N 			& Sample		\\ 
		&		& 					& (0.04~\AA\ bin)		& size		\\ \hline
346 		& blue	& 339.3 -- 339.4		& 780\tablefootmark{a}	& 204		\\
437 		& blue	& 398.8 -- 398.9		& 1070\tablefootmark{a}	& 187		\\
564		& red-L	& 511.0 -- 511.1		& 1090\tablefootmark{b}	& 205		\\
		& red-U	& 613.1 -- 613.2		& 1020\tablefootmark{b}	& 205		\\
860		& red-L	& 675.35 -- 6754.55		& 670\tablefootmark{b}	& 184		\\
		& red-U	& 869.9 -- 870.0		& 880\tablefootmark{b}	& 183		\\
\hline
\end{tabular}
\tablefoottext{a}{Reduction `A'}
\tablefoottext{b}{Reduction `B'}
\end{table}

\begin{figure}[th!]
\centering
\includegraphics[width=0.9\columnwidth,clip]{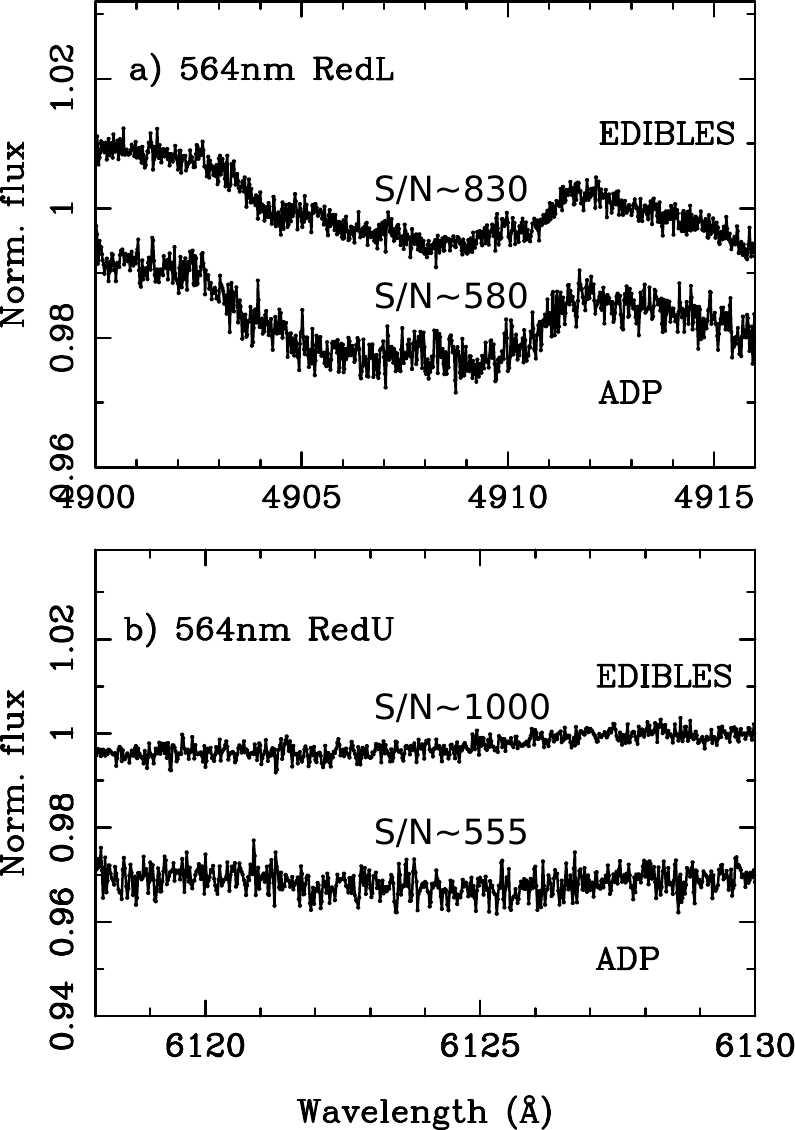}
\caption{Comparison between the ADP and EDIBLES processing of HD184915 spectra (0.02~\AA\ binsize). 
The S/N measurements, taken at 4912--4913~\AA\ and 6129--6130~\AA\ for the Red-L and Red-U spectra, are given in the top and bottom panel.
Resampling to 0.04~\AA\ binsize (i.e. corresponding to the spectral resolution) further increases the S/N by factor $\sqrt{2}$.}
\label{fig:snr}
\end{figure}

\begin{figure*}[t!]
\centering
\includegraphics[width=17.5cm]{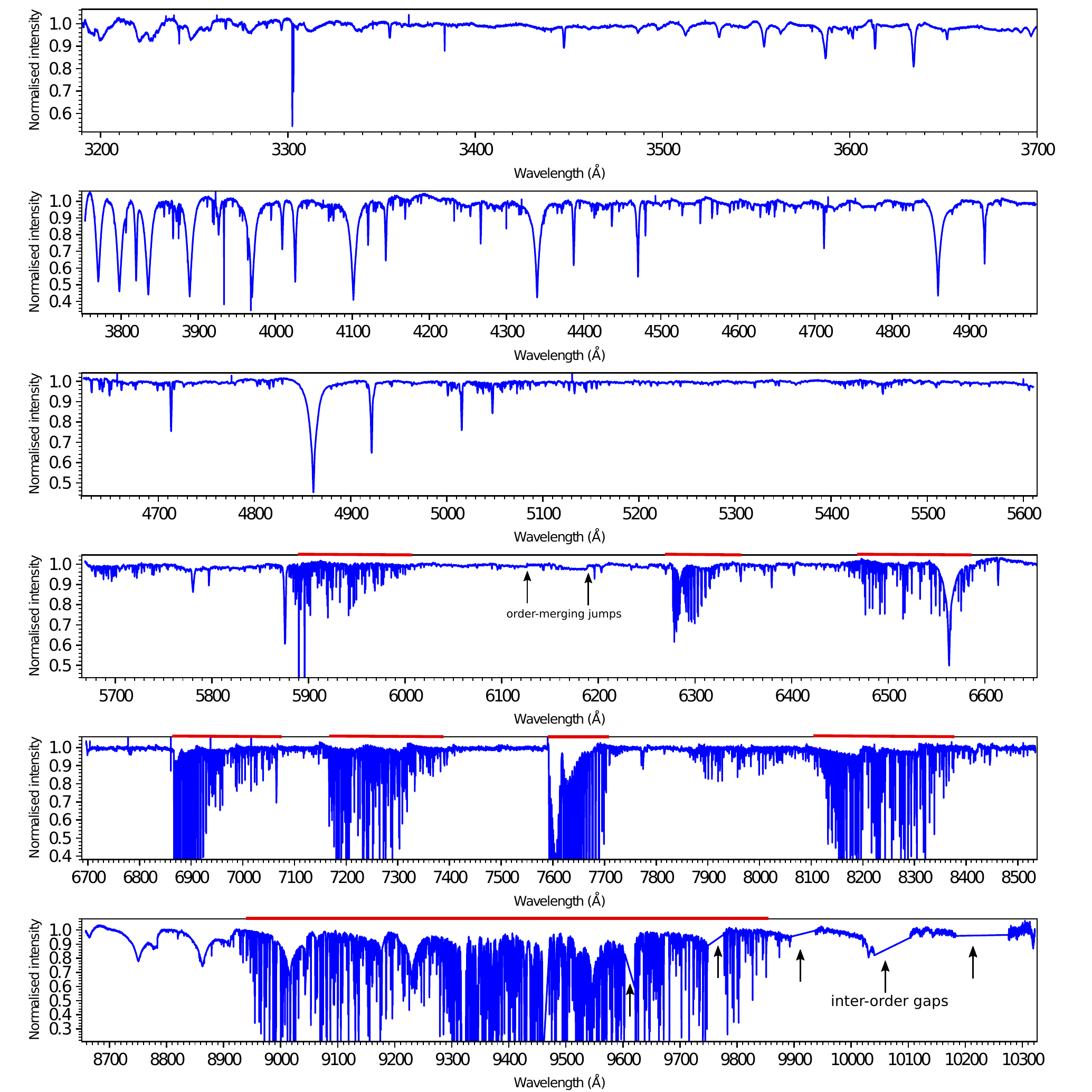}
\caption{EDIBLES UVES spectra of HD\,170740 (B2\,V) for each setting from top to bottom: 346B, 437B, 564L, 564U, 860L, 860U.
This overview figure is a demonstration of the data quality.
The main gaps in wavelength coverage are between 5610--5670~\AA\ and 8530--8680~\AA\ which correspond to the 
physical separation of the Red-L and Red-U detectors in both the 564 and 860-nm settings.
Note also the inter-order gaps, several are indicated with arrows, in the 860-nm Red-U spectrum above $\sim$9600~\AA\ as well as 
several conspicuous regions containing bands of closely-spaced telluric absorption lines (indicated with red horizontal bars) mostly in the Red-L and Red-U 860-nm spectra (bottom two panels).
Two order-merging jumps are indicated in the fourth panel.
A more detailed version of this figure is shown in the appendix (Fig.~\ref{fig:appendix:fullspec}) where specific interstellar species 
are labeled and a synthetic DIB spectrum is shown for comparison.}
\label{fig:fullspec}
\end{figure*}

\paragraph{Order extraction}

Various possible extraction methods were tested and we find that the \textit{average} extraction method with performing a flat fielding before the extraction (see above) leads to a higher S/N with respect to other methods (such as optimal extraction), which is generally the case for high S/N spectroscopy. This method also reduces the fringing seen in the red wavelengths $\geq$~700~nm, although does not entirely remove it. Fig.~\ref{fig:Plot_HD23180_860nm_Amin} shows that displays parts of an extracted science spectrum near the gap between the upper and lower CCDs. The fringing is much worse in the lower thinned EEV chip than in the upper MIT thick chip (\citealt{2009Msngr.138....8S}).

\paragraph{Cosmic ray rejection and hot pixels}

To remove the cosmic rays and possible hot pixels we applied a sigma-clipping method to each extracted spectrum. The default sigma-clipping threshold value in \texttt{esorex} is $\kappa = 10$, 
but to identify all induced hot pixels we set this $\kappa = 6$ to be sure that all cosmic/hot pixels are removed and no real data are clipped.

\paragraph{Wavelength calibration}

UVES wavelength calibration errors are typically in the range of 0.1--0.5~km\,s$^{-1}$ (\citealt{2010ApJ...723...89W}).
To achieve an accurate wavelength calibration, the dispersion relation is obtained by extracting the thorium-argon arc lamp frames using the same weights as those used for the science objects. Therefore, first we reduced the science frames with an optimal extraction method to generate a pixel weight map. Then by applying this weight map to the wavelength calibration, an accurate wavelength calibration with a statistical error less than $5 \times 10^{-4}$~\AA\ typically in all central wavelengths (for more details see Table~\ref{tb:error}) and a systematic uncertainty less than $1.7 \times 10^{-4}$~\AA\ is achieved. 
To optimize the number of lines used in the wavelength calibration solution we tested a tolerance value $\simeq 0.07$~pixels to reject the line identification with wavelength residuals worse than the tolerance. For the final iterations of the fit of the wavelength calibration solutions we set the sigma-clipping to $\kappa = 3$. 
Further improvements to the wavelength calibration are being studied for the public release of the data. For example, there is a temperature and density dependent shift (which can be as much as 1 pixel and different for the different wavelength regions) in the position of the thorium-argon lines (UVES User Manual). This can be potentially corrected by taking into account the difference between the observation and calibration temperatures and pressures. 
Also, we foresee improvements to the final wavelength calibration in the 860~nm Red-U arm, where there are few thorium-argon lines, by cross-referencing with a model telluric transmission spectrum.

\begin{figure*}[ht!]
\centering
\includegraphics[height=5.5cm]{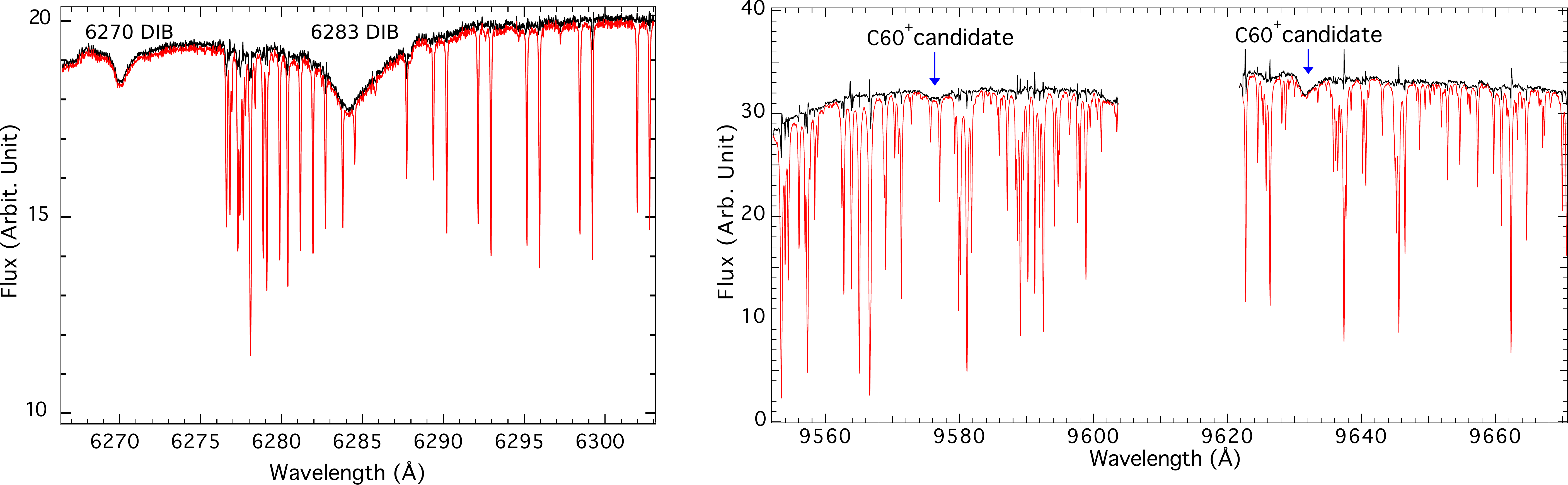}
\caption{(Left) Example of telluric line correction in the weak line regime by means of the {\it rope-length} method applied to the spectrum of HD\,170740. Telluric lines of O$_{2}$ are corrected first, then H$_{2}$O lines.
(Right) Example of telluric line correction in the strong line regime by means of a two-step method and a composite instrumental profile adjustment (see text). Residuals remain at the location of the deepest lines, especially when the model does not predict their shape and exact Doppler shift very accurately. The positions of the 9577 and 9632~\AA\ DIBs attributed to C$_{60}^+$ are indicated.}
\label{fig:6284}\label{fig:9577}
\end{figure*}

\paragraph{Blaze function and order merging}

Previous analyses of data obtained with UVES demonstrate that the shape and position of the blaze function is the primary source of problems in the order overlap regions. The continuum changes vary smoothly over subsequent orders, which may be related to the fact that the blaze profiles produced by UVES are not the same as the theoretical predictions. Accordingly, the blaze function at the overlapping regions is not sufficiently well characterised, therefore by averaging the overlap regions some artificial features appear and cause the signal level to fall off at the end of the orders.  Irregular variations in the continua at the edges of orders are likely due to a mismatch between the paths of the light from the star and the flat-field lamp (\citealt{2008eic..work..365N}) and result in discontinuities where orders have been merged. 

For reduction `B', no blaze correction was performed. Optimal merging provided the best results in the overlapping regions for both reductions `A' and `B'.
Fig.~\ref{PlotHD23180_564} shows a comparison of reductions `A' and `B' for HD\,23180 in a spectral region with two overlapping orders.
Reduction `B' reveals a mismatch of the order-overlapping regions, resulting in a jump in the spectra at 5855~\AA\ not present in the reduction `A' spectrum.

\paragraph{Quality control}

The S/N of the spectra as a function of wavelength was estimated by fitting a first order polynomial to regions of the spectra in bins of 1~\AA\  and measuring the residual in each 0.02~\AA\ wavelength bin.
Table~\ref{tb:snr} lists the median S/N (per 2-pixel ``resolution-element'') for each setting, together with the respective continuum wavelength regions.
In Fig.~\ref{fig:snr} we compare for HD\,184915 the spectra obtained with the dedicated EDIBLES processing presented here and the default archive data products (ADP) provided by ESO.
\footnote{These spectra are generated by ESO's Quality Control Group for all UVES point-source observations. 
The pipeline processing is done automatically through a dedicated workflow (with no fine-tuning of pipeline parameters specific to the needs of our programme).
The 1d-extracted spectra are ingested into the ESO archive as so-called ``Phase 3'' data products.}
The increase in S/N (labelled in the figure) is almost a factor two for both the Red-L and Red-U spectra (cf. Table~\ref{tb:setup}).
For each target the different settings/arms generally have different S/N ratios.
This is primarily due to (1) the choice of targets which are often brighter in V band compared to B band (i.e. reddening), and (2) the lower efficiency of UVES in the very blue and very red parts of the accessible wavelength range.
In terms of S/N reduction `A' performs slightly better than reduction 'B' for the 346-nm setting, both perform similar for the 437-nm and 860-nm settings, while
reduction `B' performs better for the 564-nm setting. The main difference between the two reductions is in terms of the order-merging jumps. In addition, there are small,
though noticeable differences in the `noise' between both reductions. The reduction scheme `A' is adopted as the primary scheme for results shown in this work.
Both reductions are therefore retained for reference and as a control for spurious features.

As an example, the final continuum normalised EDIBLES UVES spectrum of HD\,170740 (from reduction `A') is shown in full in Fig.~\ref{fig:fullspec}, 
where each panel corresponds to  one of the instrument settings (Table~\ref{tb:setup}). A closer view of this spectrum in shown in Appendix~\ref{appendix:fullspec}.
All processed EDIBLES spectra will be released as new ``Phase 3'' data to the ESO archive later in the project.

\section{Telluric and stellar features}\label{subsec:telluric}

\subsection{Earth transmission spectrum}

A large number of weak and strong telluric oxygen and water absorption bands arise from molecules present in the Earth atmosphere covering nearly the full DIB range.
In addition, for the near-UV spectral domain (300--350~nm) a correction for atmospheric ozone needs to be applied.

These telluric lines can be removed or modelled by using bright, early-type type star spectra recorded in the same conditions as the targets, 
or by using synthetic atmospheric transmittance spectra. 
The former method requires observing (unreddened) standard stars at roughly similar airmass, shortly before or after the primary science target observations. 
This procedure is not possible within the filler observation strategy, and as the EDIBLES targets are themselves bright this would double the required observing time.
The latter method has the advantage of saving observing time, avoiding features associated with the standard star and benefits from the increase in quality and 
availability of the molecular databases. The first tests of atmospheric spectra correction were performed based on the following approach:

First, the telluric transmittance is optimally adapted to each target. Paranal observing conditions are downloaded from the TAPAS facility\footnote{\url{http://ether.ipsl.jussieu.fr/tapas/}} (\citealt{2014A&A...564A..46B}). 
The TAPAS transmittance spectra are based on the latest HITRAN molecular database (\citealt{2013JQSRT.130....4R}), 
radiative transfer computations (Line-By-Line Radiative Transfer Model; \citealt{1995JGR...10016519C}), 
and are computed for atmospheric temperature, pressure and composition interpolated by the ETHER data centre\footnote{\url{www.pole-ether.fr}} based on a combination meteorological field observations and other information.

Then, in regions of moderate absorption (i.e. less than 70\% at line centers before instrumental broadening), a simple method called {\it rope-length} minimization is used (\citealt{2012A&A...544A.136R}). Briefly, the algorithm searches for the minimal length of the spectrum that is obtained after division of the data by the transmittance model. To do so the column of the absorbing species, the Doppler shift and the width of the instrumental function by which the transmittance model is convolved are all tuned. The method uses the fact that when the telluric lines are not well reproduced, strong maxima and minima remain after the data-model division and the spectrum length increases. On the contrary, if the modelled lines follow very well the observed ones, the corrected spectrum is smooth.  An example of correction is shown in Fig.~\ref{fig:6284} for the 6284~\AA~DIB. Here the rope-length method has been first applied to the O$_{2}$ lines, then in a subsequent step the O$_{2}$-corrected spectrum is corrected for the weak H$_{2}$O lines that are also present in this spectral region.

For regions of stronger telluric lines, division by deep lines induces overshoot and the exact shape of the telluric profiles and of the instrumental function becomes crucial. The simple rope-length method alone is no longer appropriate.  We have tested an iterative method that is a combination of the rope-length method and a classical fitting. In a first step we excluded all spectral intervals around the centres of the deep lines and performed a {\it segmented rope-length} optimization, i.e. the algorithm searches for the model parameters that minimize the sum of the lengths of the individual spectral segments. We then divided the data by the corresponding adjusted model and performed a running average of the divided spectrum to obtain an approximate stellar continuum. We then fitted the data to the convolved product of this continuum and a telluric transmittance. The instrumental function is now modelled as the sum of a Gaussian and a Lorentzian, which allows weak extended wings to be taken into account. The instrumental profile is the same for all lines within the corrected interval.
The parameters defining these two components as well as the column of the absorbing species are free to vary during the adjustment. The data were then divided by this updated model. This process can be iterated and stopped when there is no longer any decrease of the "rope-length". An example of such a two-step correction is shown in Fig.~\ref{fig:9577} for the spectral region of the C$_{60}^+$ 9577 and 9632~\AA\ DIBs. 
There are still some residuals, but these are limited mostly to the deepest, (partially) saturated telluric lines.
This occurs particularly when these lines are slightly Doppler shifted or broadened due to  atmospheric pressure in a way that is not fully predicted by the model. For a description of such effects see \citet{2014A&A...564A..46B}. 
Nevertheless, the DIBs at 9577 and 9632~\AA, assigned to be due to C$_{60}^+$ as mentioned in the introduction, stand out clearly in the telluric corrected spectrum. The three weaker DIBs between 9350 and 9450~\AA\ reported by \citet{2015ApJ...812L...8W}, though not yet confirmed independently 
(\citealt{2017MNRAS.465.3956G,2017arXiv170401501C}) present significant challenges for detection due to the presence of strong, saturated telluric water lines in this spectral range.
We intend to investigate the C$_{60}^+$ bands in more detail later in the EDIBLES project, but this will depend upon the accuracy and success of the telluric line modelling for each line-of-sight (as there are numerous saturated atmospheric water absorption lines in this wavelength region) as well as the stellar atmosphere modelling required to account for, for example, contribution of  \ion{Mg}{ii} (as discussed in \citealt{2017MNRAS.465.3956G}).

\begin{figure*}[th!]
\centering
\includegraphics[viewport=70 30 620 360,height=7.5cm,clip]{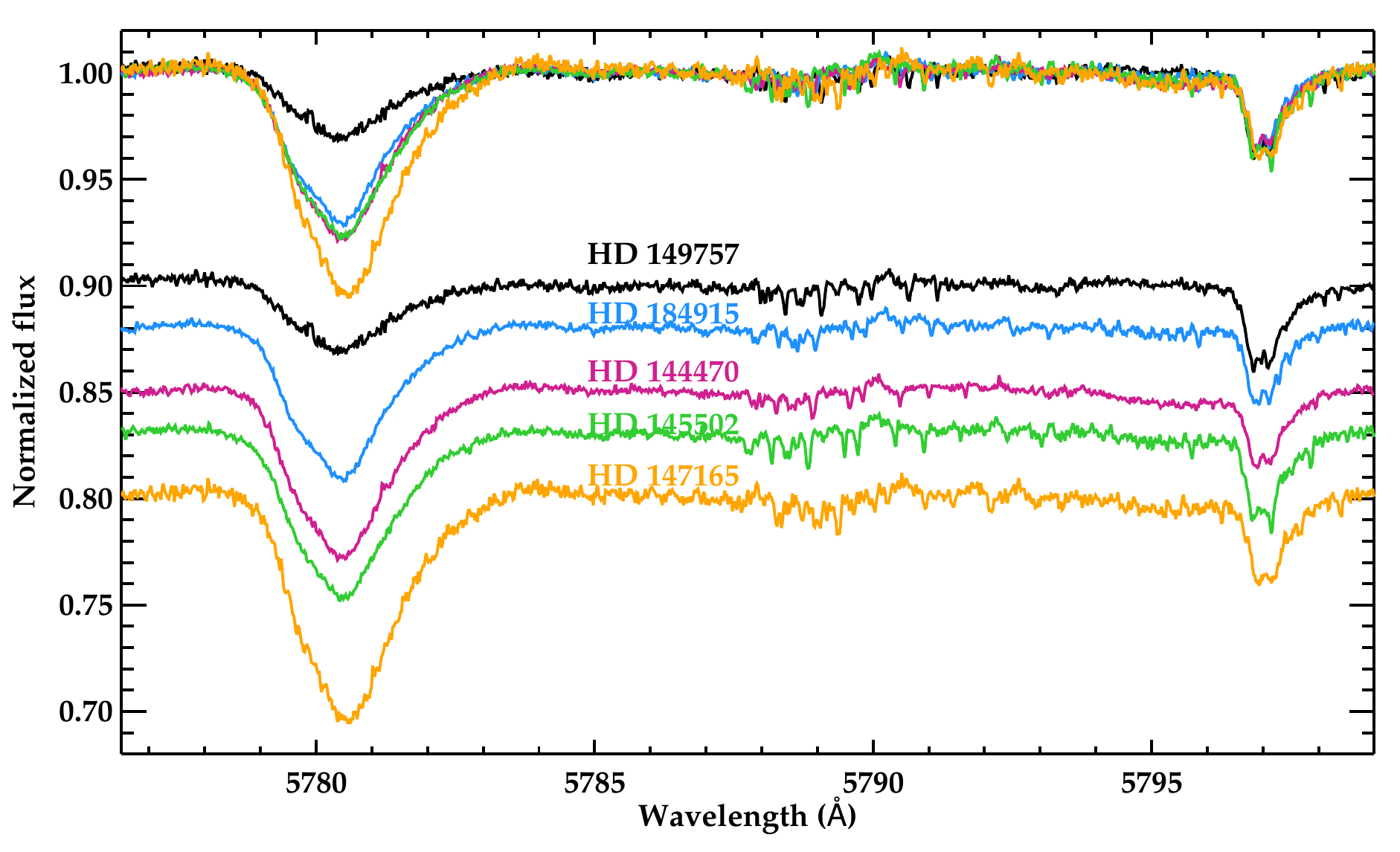}
\includegraphics[viewport=70 30 390 580,height=8.5cm,clip]{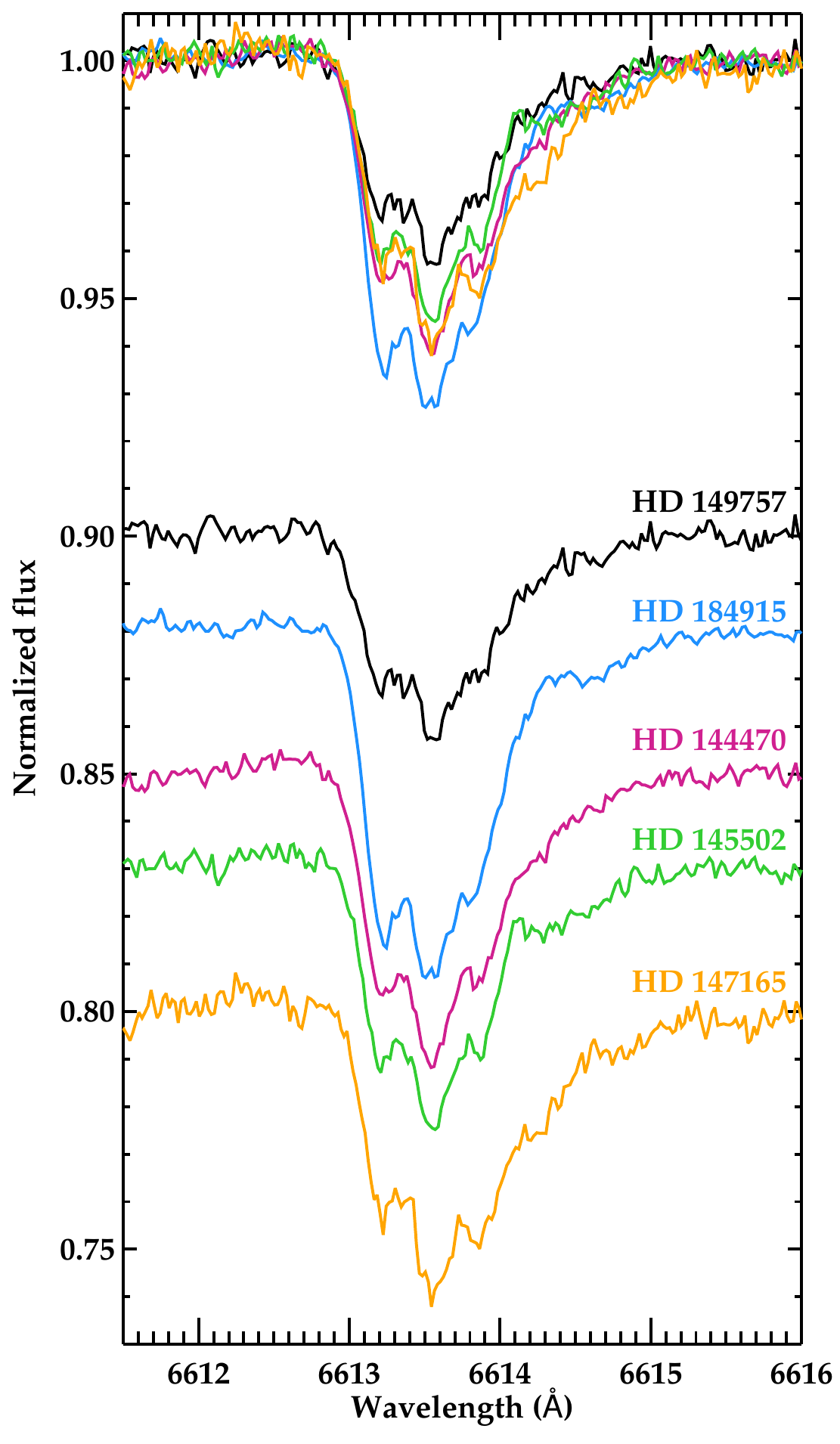}
\caption{An illustration of the quality of the spectra for the 5780 and 5797~\AA\ (left) and 6614~\AA\ (right) DIBs. 
For each target all observed spectra were co-added in the heliocentric reference frame. 
The five targets shown have comparable E(B-V) values (Tables~\ref{tb:targetlist1} and~\ref{tb:targetlist2}), and thus have similar 5797~\AA\ DIB strengths. 
The well established, strongly variable 5780/5797 ratio can be seen in the spectra, with the intensity of 5780 absorption inversely related to the molecular gas fraction, \fHtwo\ (Tables~\ref{tb:targetlist1} and~\ref{tb:targetlist2}).
The five spectra are shown superimposed on each other at the top of the panels.
Note that because of generally poor observing conditions there are numerous weak and narrow atmospheric water features present (particularly noticeable around 5788--5792~\AA) that could influence the 5797~\AA\ profile. 
In the future these features will be removed using the method described in Sect.~\ref{subsec:telluric}.}
\label{fig:edibles_sightlines}
\end{figure*}

\subsection{Stellar spectra}

In addition to the goals discussed above, the EDIBLES observations provide high-quality spectra of the target stars themselves. 
These span the full spectral range of early-type stars, from early O-type dwarfs through to late B-type supergiants (plus a couple of later/cooler stars). 
Published spectral classifications of the sample are summarised in Tables~\ref{tb:targetlist1} and \ref{tb:targetlist2}.

All of the O-type EDIBLES targets have been observed as part of the Galactic O-Star Spectroscopic Survey (GOSSS), 
a comprehensive survey of bright Galactic O stars at a resolving power of $R$\,$\sim$2500 \citep{s11,s14}. 
The detailed O-star classifications quoted in Tables~\ref{tb:targetlist1} and \ref{tb:targetlist2} are the GOSSS types -- a thorough
description of the classification criteria was given by \citet{s11}, including an overview of the various classification qualifiers 
used to convey additional information on the spectra (their Table~3).

In contrast, most of the B-type stars in the EDIBLES sample have not been subject to such morphological rigour with high-quality digital spectroscopy. 
Work is underway within the GOSSS to better define spectral standards and the classification framework for early B-type stars (Villase\~nor et al. in prep.), 
with a few stars overlapping with the EDIBLES sample. Nonetheless, the high-quality, high-resolution spectra from EDIBLES will be useful to refine 
the classification framework for B-type stars, particularly compared to similar efforts in the Magellanic Clouds \citep[e.g.][]{e15}.

In the short-term we will inspect the EDIBLES data in the context of classification, to update/refine spectral types as required -- 
whether arising from the added information of the high-resolution UVES data (cf. lower-resolution spectroscopy from the GOSSS, for example), 
or simply from intrinsic spectral variability, which is seen in many early-type stars. 
This will ensure the best parameters are adopted in estimating stellar colours, thence the line-of-sight extinctions. 
We will also look for evidence of spectroscopic binaries in targets with multiple observations (and/or relevant archival data, see, e.g., Sect.~\ref{subsec:uvespop}).
Our longer-term objective is a quantitative analysis of the stellar spectra to determine their physical parameters 
(effective temperature, gravities, rotational velocities, mass-loss rates etc), employing tools developed specifically for the analysis of early-type spectra 
\citep[e.g.][]{m05,ssd11}. 
Ultimately this will help removal of stellar features from the \mbox{EDIBLES} spectra to aid analysis of the interstellar features.

\section{Quality assessment: the interstellar spectrum}\label{sec:preview}

In this Section we highlight the interstellar lines and bands observed for a few selected lines-of-sight.

\subsection{EDIBLES}

As noted above, the full spectrum of HD\,170740 is shown as an example in Figs.~\ref{fig:fullspec} and~\ref{fig:appendix:fullspec}. 
For initial guidance in identifying the DIBs in this line-of-sight the average ISM DIB spectrum (\citealt{1994A&AS..106...39J}), scaled to \Ebv = 0.5~mag, is shown in the latter figure. 
This reference spectrum includes broad DIBs not included in e.g. \citet{2009ApJ...705...32H} but which appear to be present in the observed spectrum.

In Fig.~\ref{fig:edibles_sightlines} we compare the three strong DIBs at 5780, 5797~\AA\ (\citealt{1922LicOB..10..141H,1938ApJ....87....9M}) and 6614~\AA\ for single-cloud lines-of-sight towards 
five EDIBLES targets, HD\,149757, HD\,184915, HD\,144470, HD\,145502, and HD\,147165.  
The line of sight reddening, \Ebv, for these sightlines differs by less than 0.17~mag (c.f. Table~\ref{tb:targetlist1}). 
The spectra are averages of individual exposures co-added in the heliocentric rest frame and subsequently continuum normalised (but not otherwise scaled other than to offset them in the figure). 

As expected, for sightlines with such small variations in reddening, the  5797~\AA\ DIB profiles have similar central depths (insensitive to \fHtwo; Cami et al. 1997).
The large variations in the strength of the 5780~\AA\ DIB are thought to be related to variations in the local interstellar conditions. 
The sightline with the weakest 5780~\AA\ DIB is more neutral as indicated by the large molecular fraction, whereas the sightline with the strongest 5780~\AA\ DIB has the highest atomic column density. 
However, while the 6614~\AA\ DIB of the former is also the weakest, the latter does not exhibit the strongest 6614~\AA\ DIB, thus indicating that additional parameters must play a role in determining the DIB carrier column density.

\subsection{EDIBLES  versus CES}

\begin{figure}[h!]
\centering
\includegraphics[width=0.9\columnwidth]{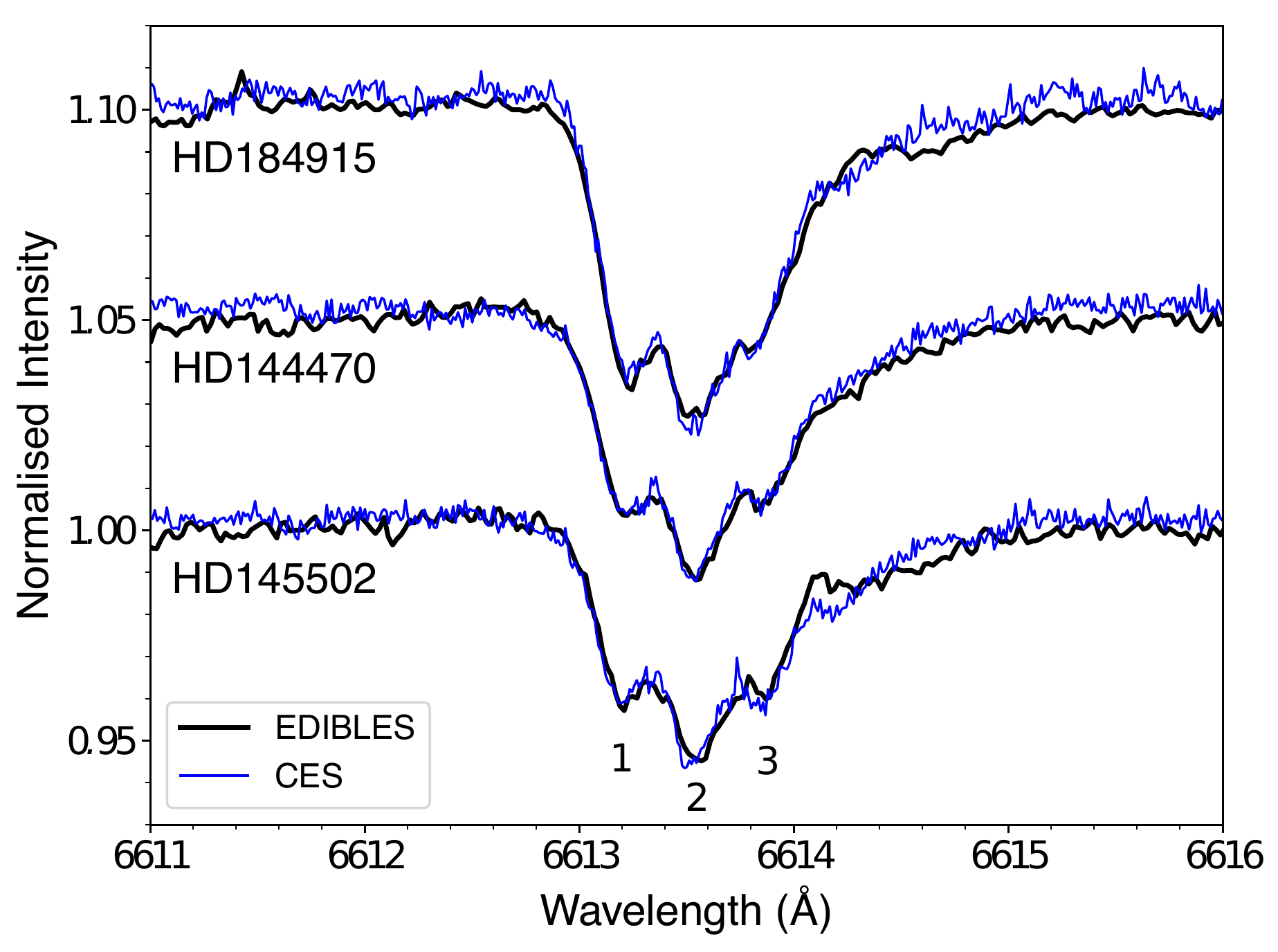}
\caption{Comparison of the 6614~\AA\ DIB for HD\,184915, HD\,144470, and HD\,145502 obtained with 
EDIBLES (black solid line; this work, $R \sim 110\,000$) and the CES (blue solid line; \mbox{\citealt{2002A&A...384..215G}}; $R \sim 220\,000$).
The sub-structure components 1, 2, and 3 are labelled in the bottom trace (see text).}
\label{fig:ces}
\end{figure}

High-resolution (R = 220,000) spectra of the 6614~\AA\ band have been recorded using the Coude \'Echelle Spectrograph (CES) 
fed by the fibre link with the Cassegrain focus of the 3.6 m telescope at La Silla Observatory (\citealt{2002A&A...384..215G}).  
A signal-to-noise of $\sim$600--1000 was achieved.  
Comparison of the EDIBLES and CES data is shown in Fig.~\ref{fig:ces} for HD\,184915, HD\,144470 and HD\,145502 and the main features are in good agreement.  
Of the three principal absorption components, the shortest wavelength feature (component 1) is relatively strong for HD\,184915 in both studies, whereas components 1 
and 3 have comparable intensities for the lines-of-sight towards HD\,144470 and HD\,145502.  
No significant additional structure is evident in the higher resolution CES spectra.

\begin{figure}[h!]
\centering
\includegraphics[width=0.9\columnwidth]{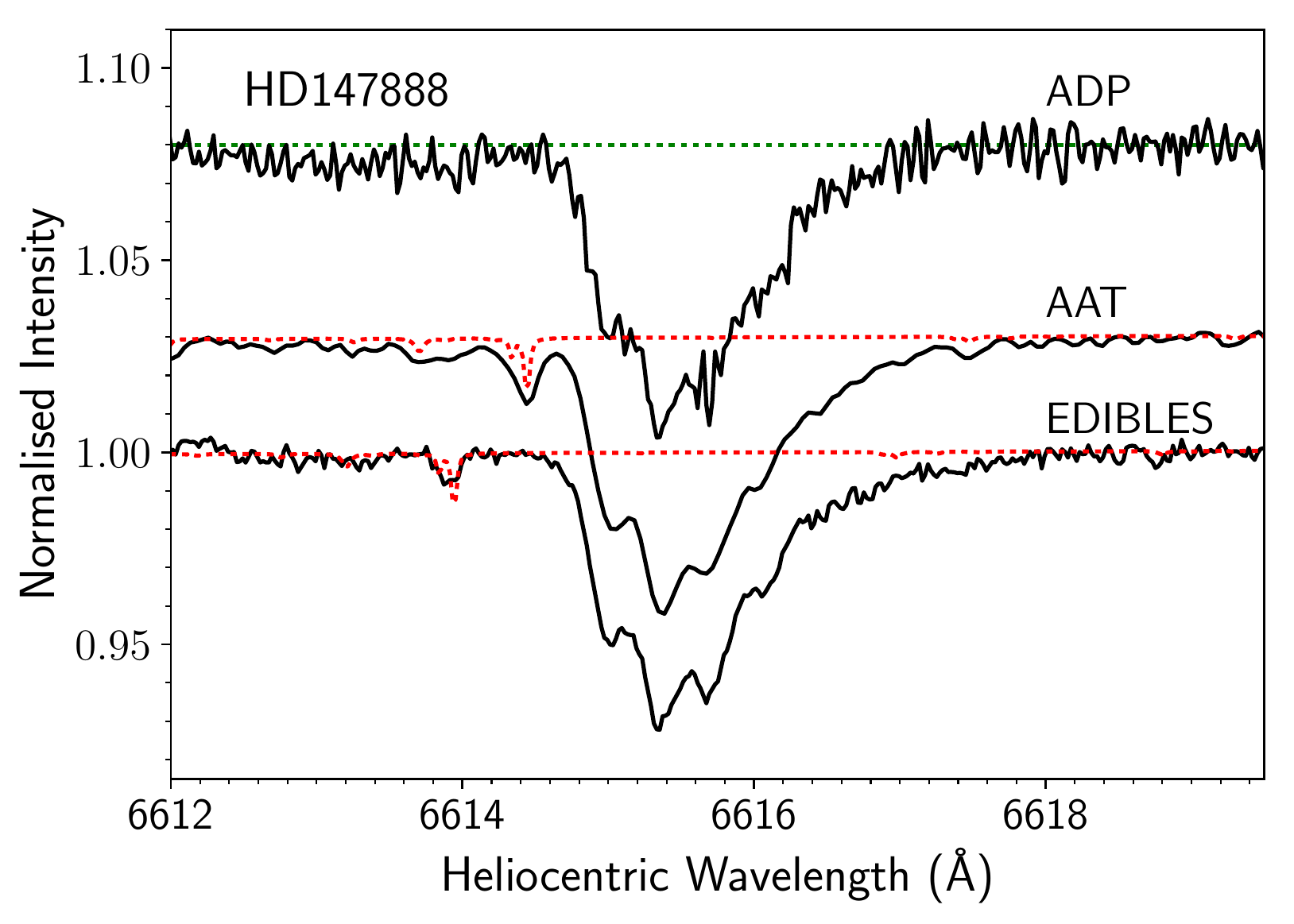}
\caption{Comparison of the 6614~\AA\ DIB for HD\,147888 ($\rho$ Oph D) obtained with EDIBLES (this work, $R \sim 110\,000$) and the AAT (\citealt{2013ApJ...764L..10C}, $R \sim 58\,000$). 
The top spectrum labeled ``ADP'' is the spectrum obtained with the standard ESO archive pipeline processing (i.e. using the default number of 5 flat-field frames). 
The red dotted line represents the telluric absorption spectrum, shifted to match the heliocentric rest frame of each target; highlighting the presence of a small telluric absorption feature at 6614/6614.5~\AA.}
\label{fig:edibles_aat}
\end{figure}

\begin{figure*}[!ht]
    \begin{minipage}[t][][b]{.33\textwidth}
        \centering
 		\includegraphics[height=5.cm]{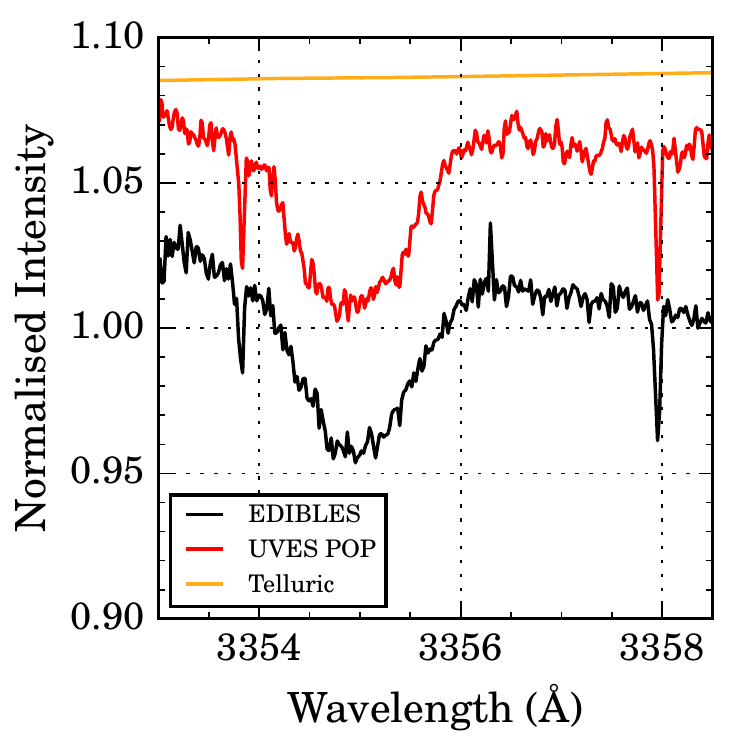}
 		\includegraphics[height=5.cm]{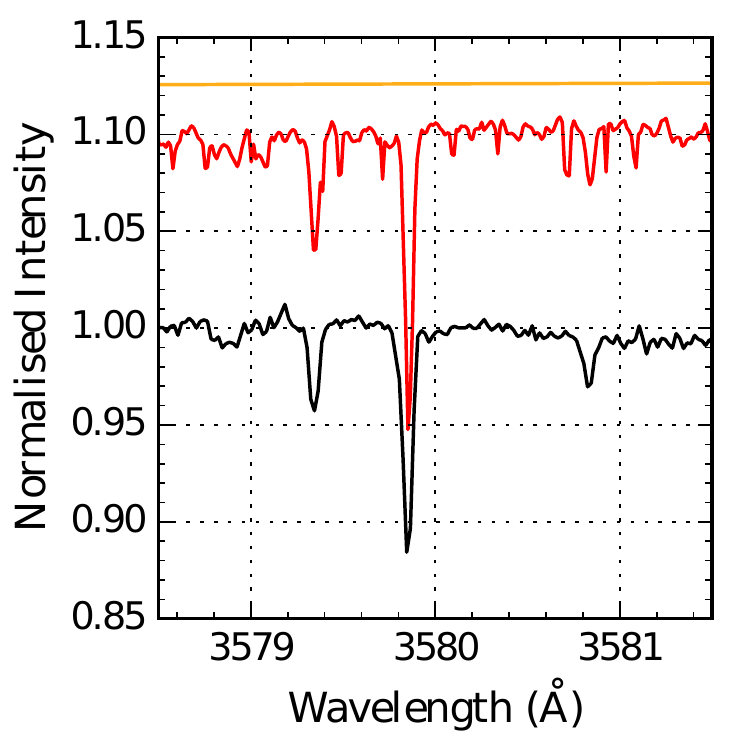} 
       \caption{Comparison of EDIBLES (black) and UVES POP (red) spectra of HD\,169454 for interstellar lines of
       NH($\lambda\lambda$3353.92, 3358.05~\AA) (top) 
       and CN(1-0) ($\lambda\lambda$3579.45, 3579.96, and 3580.9~\AA; \citealt{1989ApJ...343L...1M}) (bottom);
       the telluric spectrum is shown in orange. 
	HD\,169454 is a blue supergiant (B1 Ia) -- the broad stellar line in the spectrum is \ion{He}{i} $\lambda$3355~\AA.}
       \label{fig:edibles_uvespop1}
    \end{minipage}%
    \hspace{5mm}
    \begin{minipage}[t][][b]{0.63\textwidth}
        \centering
		\includegraphics[height=5.cm]{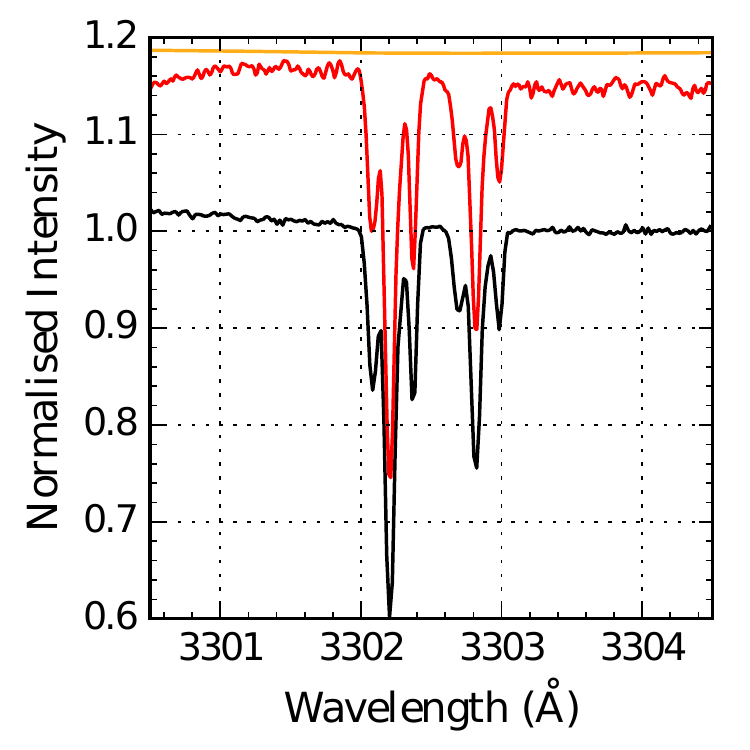}
		\includegraphics[height=5.cm]{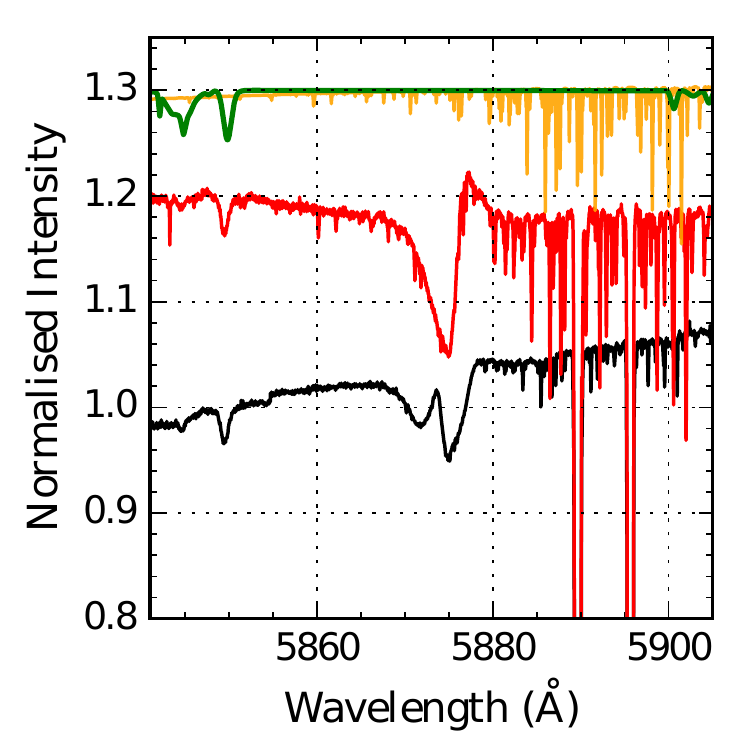} \\
		\includegraphics[height=5.cm]{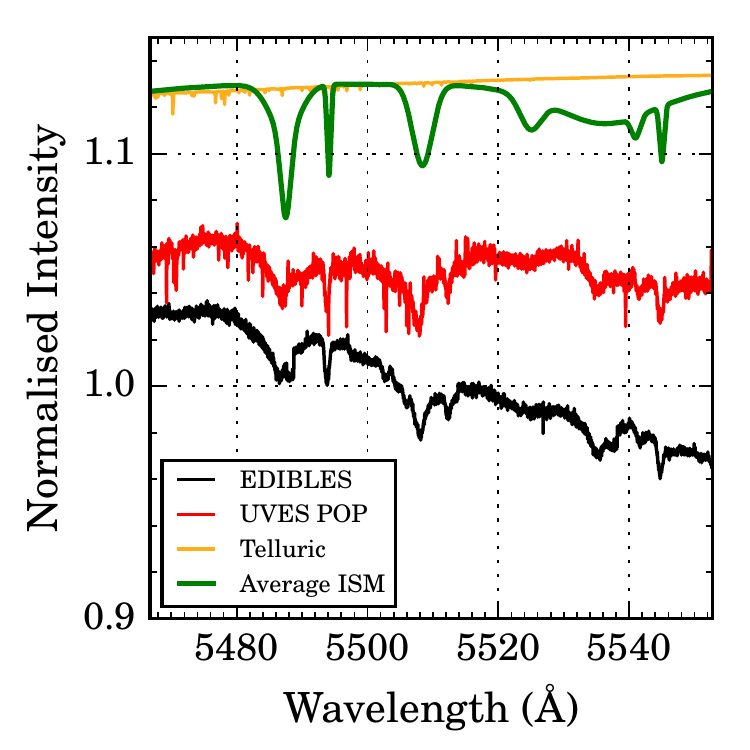}
		\includegraphics[height=5.cm]{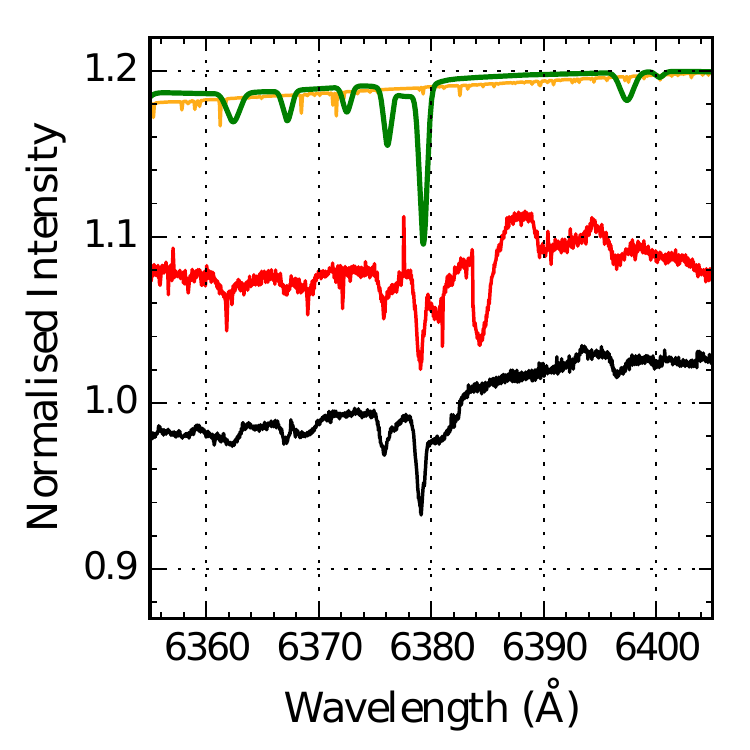}
      \caption{Comparison of EDIBLES (black) and UVES POP (red) spectra of HD\,148937 (O6 f?p) for the interstellar Na lines 
      	(UV, top left; D, top right) and DIBs at $\lambda\lambda$5480--5545\,\AA\ (lower left) and $\lambda\lambda$6360--6379\,\AA\ (lower right).
	The telluric spectrum is shown in orange, and the average ISM DIB spectrum in green.
	The apparent feature at 6384\,\AA\ in the UVES POP data (bottom right panel) is related to the order merging, 
	but the significant change in the \ion{He}{i}\,5876 line (top right panel) appears astrophysical in nature.}
        \label{fig:edibles_uvespop2}
    \end{minipage}
\end{figure*}

\subsection{EDIBLES versus AAT}

To assess the quality of our spectra with respect to previous studies, we compared our EDIBLES spectrum of HD147888 (obtained in a single exposure on 2016-08-08) with a spectrum obtained using the Anglo-Australian Telescope (AAT) in June 2004 by \citet{2013ApJ...764L..10C}. 
The AAT spectra were obtained using the UCLES instrument at a resolving power of 58\,000 and have S/N~$\sim$~900. \citet{2013ApJ...764L..10C} reduced their data using a custom procedure, taking special care to optimally correct for CCD non-linearity, scattered light subtraction and flat fielding, as well as wavelength calibration. 
A comparison between the EDIBLES and AAT spectra is shown in Fig.~\ref{fig:edibles_aat} for the 6614~\AA\ DIB, which makes for a good general test case due to the presence of narrow and broad features within its profile. 
Apart from differences in the wavelengths of the telluric features, and a slight difference in the overall spectral slope (presumably due to uncorrected differences between the UVES and UCLES blaze functions), the spectra are almost identical within the noise.  Slight differences in the depths and widths of the substructure peaks are attributable to the higher resolution of the UVES spectrum. 
The overall excellent match demonstrates the quality of our EDIBLES reduction procedure. For reference we show also the ADP spectrum, which uses only 5 flat-field frames; the large increase in S/N when using our custom flat-field processing is apparent.

\subsection{EDIBLES versus UVES POP}\label{subsec:uvespop}

To further illustrate the quality of the data we compared spectra of two targets, HD\,148937 (Fig.~\ref{fig:edibles_uvespop1}) and HD\,169454 (Fig.~\ref{fig:edibles_uvespop2}), which were both observed as part of the EDIBLES survey and the UVES Paranal Observatory Project (POP; \citealt{2003Msngr.114...10B}).
The UVES POP programme gathered a library of high-resolution, high S/N spectra of (field) stars across the Hertzsprung-Russell diagram.
The spectra in the figures have been scaled in intensity to facilitate comparison, but not continuum normalised.

Fig.~\ref{fig:edibles_uvespop1} shows the NH and OH$^+$ lines in the sightline towards HD\,169454.
The top panels in Fig.~\ref{fig:edibles_uvespop2} compare
the sodium doublets at $\sim$3303~\AA\ (see also \citealt{2006MNRAS.367.1478H}) and $\sim$5895~\AA\ for the sightline towards HD\,148937.
The bottom panel compares several weak and strong DIBs in the HD\,148937 line-of-sight.
In each panel the EDIBLES and UVES POP spectra are shown in black and red, respectively. The average ISM synthetic DIB spectrum (adapted from \citealt{1994A&AS..106...39J}) is shown in green and a generic Paranal model telluric spectrum\footnote{generated from the ESO SkyCalc Sky Model Calculator} is shown in orange. 
The agreement between stellar and interstellar features in the EDIBLES and UVES POP is excellent, where the EDIBLES data generally reach a higher S/N.

The two right-hand panels of Fig.~\ref{fig:edibles_uvespop2} reveal apparent variations between the EDIBLES data compared to the UVES POP spectrum. The strong absorption in the POP data at $\sim$6384~\AA\ coincides with the order ends/overlap, and appears to be an artefact of the order merging. In contrast, the significant change in the He\,I 5876 absorption seems robust. In the context of HD\,148937 being a peculiar magnetic star, this variation is quite remarkable. This aspect will be discussed elsewhere.

These comparisons with existing data for selected sightlines show the excellent data quality of the spectra acquired within EDIBLES, illustrating its potential for detailed studies of physical conditions and DIB properties in the diffuse ISM.

\section{Summary}\label{sec:summary}

In this first of a series of papers we have presented the design and scope of the ESO Diffuse Interstellar Bands Large Exploration Survey (EDIBLES). 
We presented the scientific goals and the immediate objectives of EDIBLES, along with the survey sample and its characteristics.

At the time of writing (May 2017), spectra had been acquired for 96 targets from 114 in the overall programme. 
These spectra cover the wavelength range from 305 to 1042~nm at a spectral resolving power of $\sim$70\,000--110\,000. 
We have presented the data-processing steps  employed to reduce the survey data so far.

The median S/N ratio (per 0.04~\AA\ spectral bin) varies from $\sim$600--700 hundred in the blue ($<$400 nm) and near-infrared ($>$800 nm) ranges, 
to $\geq$1000 in the green-red (500--700 nm). 
To illustrate the quality and scope of the new spectra we have compared 
(1) EDIBLES and AAT spectra of the 6613~\AA\ DIB towards HD\,147888, 
(2) EDIBLES and CES spectra of the 6613~\AA\ DIB for the sightlines towards HD\,184915, HD\,144470 and HD\,145502, and 
(3) EDIBLES and UVES POP spectra of HD\,148937 and HD\,169454.

Upcoming papers in this series will present in detail the array of scientific results that are being explored with the \mbox{EDIBLES} data set.
Once the program is completed the advanced data products (merged and normalised spectra) will be released to the community through the ESO Science Archive and the CDS/ViZieR service.

\begin{acknowledgements}
This work is based on observations collected at the European Organisation for Astronomical Research in the Southern Hemisphere under ESO programmes 194.C-0833 and 266.D-5655.
The EDIBLES project was initiated at the IAU Symposium 297 "The Diffuse Interstellar Bands" \citep{2014IAUS..297.....C}; the support of the International Astronomical Union for this meeting is gratefully acknowledged. 
NLJC thanks the Paranal Observatory staff and the ESO User Support Department for their assistance with successfully executing the observations.
The authors thank both the Royal Astronomical Society and the Lorentz Center for hosting workshops, and Jacek Kre{\l}owski for making available the CES (ESO) data.
We thank the referee for constructive comments that helped improve the paper.
The research leading to these results has received funding from the European Research Council under the European Union's Seventh Framework Programme 
(FP/2007-2013) ERC-2013-SyG, Grant Agreement n. 610256 NANOCOSMOS. 
FS acknowledges the support of NASA through the APRA SMD Program.
AMI acknowledges support from Agence Nationale de la Recherche through the STILISM project (ANR-12-BS05-0016-02) and from the Spanish PNAYA through project \mbox{AYA2015-68217-P}.
\end{acknowledgements}

\bibliographystyle{aa}  
\bibliography{edibles} %


\onecolumn

\appendix

\section{Survey sample}\label{app:targetlist}

Table~\ref{tb:targetlist1} lists spectral classifications and line-of-sight parameters for the observed EDIBLES targets to date; Table~\ref{tb:targetlist2} lists the additional observations that will
supplement the data discussed here in due course. The $R_V$ and $A_V$ extinction values are taken (in order of preference) from \citet{2004ApJ...616..912V}, \citet{2007ApJ...663..320F}, or
\citet{2003AN....324..219W}. \ion{H}{i} and H$_2$ column densities are from 
\citet{2009ApJ...700.1299J}. The interstellar reddening, \Ebv, was computed from $(B-V)_J$ colours (taking the average from Tycho-2 and Simbad $B-V$ colours; where Tycho-2 colours were converted to Johnson colours; \citealt{2002AJ....124.1670M}) and the published spectral classifications, adopting intrinsic colours from \citet{1970A&A.....4..234F}. 

References for the spectral classifications in col. 4 of both tables are: M50 \citep{m50}; S52 \citep{s52}; M55 \citep{mcw55}; H56 \citep{h56}; S56 \citep[][classification originally from W.~W.~Morgan]{s56}; F57 \citep{f57}; O59 \citep{o59}; G68 \citep{g68}; L68 \citep{l68}; A69 \citep{am69}; C69 \citep{ccjj69}; H69 \citep{h69}; S71 \citep{sc71}; W71 \citep{w71}; H73 \citep{h73}; M73 \citep{mk73}; L75 \citep{l75}; L76 \citep{la76}; W76 \citep{w76}; G77 \citep{g77}; H78 \citep{h78}; H82 \citep{h82}; G94 \citep{gg94}; M95 \citep{m95}; P98 \citep{pa98}; S99 \citep{s99}; G01 \citep{G01}; E05 \citep{e05}; L06 \citep{l06}; R09 \citep{rm09}; S11 \citep{s11}; H13 \citep{h13}; S14 \citep{s14}.

\begin{landscape}
{
\setlength\LTcapwidth{\linewidth}
\begin{longtable}{lllllllllllll}
\caption{Targets observed for EDIBLES.\label{tb:targetlist1}}\\ 
\hline\hline
Identifier 		&  RA  		&  Dec  		&  Spectral type & Ref.  &  \Ebv  &  $R_V$  &  $A_V$  & Ref  	&  log $N$(\ion{H}{i}) &  log $N$(H$_2$)	&  $f_{H_2}$  &  \\
	 			& J2000		& J2000 		&		& &  mag	& 		& mag	& 			 	& cm$^{-2}$	& cm$^{-2}$	& 		&		\\ 
\hline
\endfirsthead
\caption[]{\it{continued}} \\
\hline
Identifier		&  RA  		&  Dec  		&  Spectral type & Ref.  &  \Ebv  &  $R_V$  &  $A_V$  & Ref  	&  log N(\ion{H}{i}) &  log N(H$_2$)	&  $f_{H_2}$  &  \\
	 			& J2000		& J2000 		&		& &  mag	& 		& mag	& 			 	& cm$^{-2}$	& cm$^{-2}$	& 		&		\\ 
\hline
\endhead
\hline
\multicolumn{13}{r}{\it{continued on next page}}\\
\endfoot
\hline
\endlastfoot
\object{HD 22951}		&  03:42:22.7  &  $+$33:57:54.1  &  B0.5 V  & M55 & 0.23  &  --  &  --  &    &  21.04  &  20.46  &  0.35  &  \\
\object{HD 23016}  	&  03:42:19.0  &  $+$19:42:00.9  &  B8 Ve  &  S99 & 0.05  &  2.30  &  0.12  &  W03  &  --  &  --  & --   & \\ 
\object{HD 23180}  	&  03:44:19.1  &  $+$32:17:17.7  &  B1 III  & G77 & 0.28  &  3.11  &  0.93  &  V04  & 20.88  &  20.60  &  0.51  &  \\
\object{HD 24398}  	&  03:54:07.9  &  $+$31:53:01.1  &  B1 Ib   &  L68 & 0.29  &  2.63  &  0.89  &  W03  &  20.8  &  20.67  &  0.60  &  \\
\object{HD 27778}  	&  04:23:59.8  &  $+$24:18:03.6  &  B3 V    &  O59 & 0.36  &  2.59  &  0.91 	&  V04  &  20.35  &  20.72  &  0.82  &  \\
\object{HD 34748}	&  05:19:35.3  &  $-$01:24:42.8  &  B1.5 Vn & L68 & 0.13  &  2.61  &  0.29  &  W03  &  21.16  &  --  &    &  \\
\object{HD 36486}  	&  05:32:00.4  &  $-$00:17:56.7  &  O9.5 II Nwk & S11 & 0.16  &  --    &  --  	&    	&  20.19  &  14.68  &  0.00  & \\ 
\object{HD 36695}  	&  05:33:31.5  &  $-$01:09:21.9  &  B1 V  	 &  L68 & 0.08  &  2.45  &  0.15  &  W03  &  20.7  &  --  & --   &  \\
\object{HD 36822}  	&  05:34:49.2  &  $+$09:29:22.5  &  B0 III  &  L75 & 0.14  &  2.64  &  0.21  &  W03  &  20.81  &  19.32  &  0.06  &  \\ 
\object{HD 36861}  	&  05:35:08.3  &  $+$09:56:03.0  &  O8 III ((f)) &  S11 & 0.13  &  2.46  &  0.22  &  W03  &  20.78  &  19.12  &  0.04  &  \\
\object{HD 37022}  	&  05:35:16.5  &  $-$05:23:23.2  & O7 Vp & S11 & 0.31  &  5.73  &  1.78  &  V04  &  21.54  &  17.55  & 0.00   &\\
\object{HD 37041}  	&  05:35:22.9  &  $-$05:24:57.8  & O9.5 IVp & S11 & 0.21	  &  5.67  &  1.13  &  V04  & --   & --   &  --  &  \\
\object{HD 37061}  	&  05:35:31.4  &  $-$05:16:02.6  &  B0.5 V  & S71 & 0.52 &  4.29  &  2.40  &  V04  &--  &  --  & --   &  \\
\object{HD 37367}  	&  05:39:18.3  &  $+$29:12:54.8  &  B2 IV-V & L68 & 0.37  &  3.55  &  1.49  &  V04  &  21.17  &  20.53  &  0.31  &  \\
\object{HD 37903}  	&  05:41:38.4  &  $-$02:15:32.5  &  B1.5 V &  S52 & 0.34	 &  3.74  &  1.31  &  V04  & 21.12  &  20.85  &  0.60  &  \\
\object{HD 38771}  	&  05:47:45.4  &  $-$09:40:10.6  &  B0.5 Ia & W71 & 0.07  &  --    &  --	   &       &  20.52  &  15.68  &  0.00  &  \\
\object{HD 40111}  	&  05:57:59.7  &  $+$25:57:14.1  &  B1 Ib  & L68 & 0.11  &  3.05  &  0.37  &  W03  &  20.9  &  19.73  &  0.12  &  \\
\object{HD 41117}  	&  06:03:55.2  &  $+$20:08:18.4 &  B2 Ia & W71 & 0.41  &  3.12  &  1.25  &  V04  & 20.15  &  20.69  &  0.87  &  \\
\object{HD 43384}  	&  06:16:58.7  &  $+$23:44:27.3  &  B3 Iab  & L68 & 0.54  &  3.29  &  1.91  &  W03  &  21.27  &  -- &  --  &  \\
\object{HD 53975}  	&  07:06:36.0  &  $-$12:23:38.0  &  O7.5 Vz  & S11 & 0.20  &  3.25  &  0.62  &  W03  &  21.13  &  19.18  &  0.02  &  \\
\object{HD 54439}  	&  07:08:23.2   &  $-$11:51:08.6  &  B2 IIIn & M55 &  0.27  &  2.73  &  0.79  &  V04  &  --  &  --  & --   &  \\
\object{HD 54662}  	&  07:09:20.3  &  $-$10:20:47.8  &  O7 Vz var? (SB2) & S14 & 0.32  &  3.12  &  0.81  &  W03  &  21.38  &  20.00  &  0.08  &  \\
\object{HD 55879}  	&  07:14:28.3  &  $-$10:18:58.5  &  O9.7 III & S11 & 0.13&  2.47  &  0.15  &  W03  &  20.9  &  --  &  --  &  \\
\object{HD 57061}  	&  07:18:42.5  &  $-$24:57:15.8  &  O9 II (SB2+SBE) & S14 & 0.12  &  3.57  &  0.36  &  W03  &  20.7  &  15.48  &  0.00  & \\ 
\object{HD 66194}  	&  07:58:50.6  &  $-$60:49:28.1  &  B2 Vne & A69 & 0.15  &  2.25  &  0.41  &  W03  &  --  &  --  & --   &  \\
\object{HD 66811}  	&  08:03:35.1  &  $-$40:00:11.3  &  O4 I(n)fp & S14 & 0.08  &  5.01  &  0.25  &  W03  &  19.96  &  14.45  &  0.01  &  \\
\object{HD 73882}  	&  08:39:09.5  &  $-$40:25:09.3  &  O8.5 IV (SB1+SBE) & S14 & 0.68 &  3.56  &  2.46  &  V04  &  21.11  &  21.08  &  0.65  &  \\
\object{HD 75309}  	&  08:47:28.0  &  $-$46:27:04.0  &  B2 Ib/II & H78 & 0.16  &  3.53  & 1.02   &  V04  & 21.07  &  20.20  &  0.21  &  \\
\object{HD 75860}  	&  08:50:53.2  &  $-$43:45:05.4  &  BC2 Iab  & W76 &  0.87  &  3.44  &  3.10  &  W03  &  --  &  --  & --   &  \\
\object{HD 79186}  	&  09:11:04.4  &  $-$44:52:04.4  &  B5 Ia  & H69 & 0.28  &  3.21  &  1.28  &  V04  &  21.18  &  20.72  &  0.41 &  \\
\object{HD 80558}  	&  09:18:42.4  &  $-$51:33:38.3  &  B6 Ia  & H69 & 0.57  &  3.35  &  2.01  &  W03  &  --  &  --  & --   &  \\
\object{HD 81188}	& 09:22:06.8	& $-$55:00:38.4	   & B2 IV	 & L75 & 0.10 & --	& --	&	& 19.95	& 17.70	& 0.01 &	\\ 
\object{HD 91824}  	&  10:34:46.6  &  $-$58:09:22.0  &  O7 V((f))z (SB1) & S14 & 0.25  &  3.35  &  0.80  &  V04  & 21.12  &  19.85  &  0.10  &  \\
\object{HD 93030}  	&  10:42:57.4  &  $-$64:23:40.0 &  B0.2 Vp  & W76 & 0.08  &  2.43  &  0.10  &  W03  &  20.26  &  15.02  &  0.00  & \\ 
\object{HD 93205} 	&  10:44:33.7  &  $-$59:44:15.5 &  O3.5 V((f)) + O8 V & S14 & 0.34 &  3.25  &  1.24  &  V04  &  21.38  &  19.78  &  0.05  &  \\
\object{HD 93222}  	&  10:44:36.2  &  $-$60:05:29.0  &  O7 V((f))z  & S14 & 0.34 &  4.76  &  1.71  &  V04  & 21.4  &  19.78  &  0.05  &  \\
\object{HD 93843}  	&  10:48:37.8  &  $-$60:13:25.5  &  O5 III(fc) SB1?  & S14 & 0.27 &  3.89  &  0.66  &  V04  & 21.33  &  19.61  &  0.04  &\\
\object{HD 94493}  	&  10:53:15.1  &  $-$60:48:53.2 &  B0.5 Ib: & H73 & 0.23  &  3.57  &  0.82  &  V04  &  21.08  &  20.15  &  0.19  &  \\
\object{HD 99953}  	&  11:29:15.1  &  $-$63:33:14.2  &  B2 Ia & M55 & 0.45 &  3.69  &  1.77  &  V04  &  --  & --   & --   &  \\
\object{HD 103779}  	&  11:56:58.0  &  $-$63:14:56.7 &  B0.5 Iab  & G77 &  0.24  &  3.29  &  0.69  &  V04  &  21.16  &  19.82  &  0.08  &  \\
\object{HD 104705}  	&  12:03:23.9  &  $-$62:41:45.8 &  B0.5 III & M55 & 0.23 &  2.81  &  0.65  &  V04  &  21.1  &  20.08  &  0.16  &  \\
\object{HD 109399}  	&  12:35:16.5  &  $-$72:43:00.8 &  B0.5 Ib & H73 &  0.26  &  3.26  &  0.68  &  W03  &  21.11  &  20.04  &  0.15  &  \\
\object{HD 111934}  	&  12:53:37.6  &  $-$60:21:25.4  &  B1.5 Ib  & E05 &  0.34  &  2.45  & 1.25   & V04   & --   &  --  & --   &  \\
\object{HD 112272}  	&  12:56:33.7  &  $-$64:21:39.2  &  B0.5 Ia  & M55 &  0.85  &  3.09  & 3.09   &  V04  &  --  &  --  & --    &  \\
\object{HD 113904}  	&  13:08:07.2  &  $-$65:18:21.7 &  O9 III  & S14 & 0.20  &  3.63  &  0.76  &  W03  &  21.08  &  19.83  &  0.10 &  \\ 
\object{HD 116852}  	&  13:30:23.5  &  $-$78:51:20.5 &  O8.5 II-III((f)) & S14 &   0.20  &  2.42  &  0.51  &  V04  &  20.96  &  19.79  & 0.12    &  \\
\object{HD 122879}  	&  14:06:25.2  &  $-$59:42:57.3 &  B0 Ia & W76 & 0.34  &  3.15  &  1.13  &  V04  & 21.26  &  20.24  &  0.16 &  \\
\object{HD 124314} 	&  14:15:01.6  &  $-$61:42:24.4  &  O6 IV(n)((f)) & S14 &  0.49  &  2.97  &  1.49  &  W03  &  21.41  &  20.52  &  0.21  &  \\
\object{HD 133518}  	&  15:06:56.0  &  $-$52:01:47.2&  B2 Vp  & G77 & 0.13  &  1.79  &  0.27  &  W03  &  --  &  --  & --   &  \\
\object{HD 135591}  	&  15:18:49.1  &  $-$60:18:51.5&  O8 IV((f)) & S14 & 0.20  &  3.57  &  0.79  &  W03  &  21.08  &  19.77  & 0.09   &  \\
\object{HD 143275}  	&  16:00:20.0  &  $-$22:37:18.1&  B0.3 IV & M73 &  0.19  &  3.09  &  0.43  &  W03  &  21.15  &  19.41  &  0.04  &  \\
\object{HD 144470}  	&  16:06:48.4  &  $-$20:40:09.1&  B1 V 	&  W71 & 0.21  &  3.37  &  0.74  &  V04  & 21.17  &  20.05  & 0.13   &  \\  
\object{HD 145502}  	&  16:11:59.7  & $-$19:27:38.5&  B2 V (SB2) & L75 & 0.27  & --   & --   &    &  21.07  &  19.89  & 0.12  &  \\
\object{HD 147084}  	&  16:20:38.2  & $-$24:10:09.8 &  A5 II  & G01 & 0.78   & --   & --   &    &  --  & --   & --   &  \\
\object{HD 147165}  	&  16:21:11.3  & $-$25:35:34.1&  B1 III  & W71 & 0.37  &  3.86  &  1.47  &  V04  & 21.17  &  20.05  & 0.13   &  \\
\object{HD 147683}  	&  16 24 43.7  &  $-$34 53 37.5 &  B3: Vn (SB2)  & G77 &  0.29  & --   & --   &    &  21.20   & 20.68   & 0.38   &  \\
\object{HD 147888}  	& 16:25:24.3  &  $-$23:27:36.8&  B3 V:  & F57 & 0.44  &  3.89  &  1.98  &  V04  & 21.69  &  20.48  & 0.11   &  \\
\object{HD 147889}  	&  16:25:24.3  &  $-$24:21:07.1& B2 V & H56 & 1.03 &  3.95  &  4.35  &  V04  & --   & --   & --   &  \\ 
\object{HD 147933}  	&  16:25:35.1  &  $-$23:26:48.7  & B2 IV   & H69 & 0.43   & 5.74   & 2.58   & V04   &  21.64   & 20.57 &  0.15 &  \\
\object{HD 147934}	&  16:25:35.0	&  $-$23:26:46.0  & B2 V   & H69 &  0.43   &   -- 	&    --	 &    	 &    --  &  --  & --   &  \\
\object{HD 148605}  	&  16:30:12.5 &  $-$25:06:54.8&  B2 V  & M73 & 0.07  &  3.18  &  0.25 &  W03  &  20.68  &  18.74  &  0.02  &  \\
\object{HD 148937}  	&  16:33:52.4  &  $-$48:06:40.5&  O6 f?p & S14 & 0.65  &  3.28  &  2.20  &  W03  &  21.60  &  20.71  & 0.21   &  \\
\object{HD 149038}  	&  16:34:05.0  &  $-$44:02:43.1&  O9.7 Iab & S14 & 0.31	 &  4.92  &  1.08  & V04   & 21.00  &  20.45  &  0.36  &  \\
\object{HD 149404}	&  16:36:22.6  &  $-$42:51:31.9  & O8.5 Iab(f)p (SB2) & S14 & 0.67	& 3.53	& 2.19	& V04	&  --  &  --  & --   &  \\
\object{HD 149757}  	&  16:37:09.5   & $-$10:34:01.5 & O9.2 IVnn	& S14 &  0.32 &  2.55 & 0.82  &  V04  & 20.72  &  20.65  & 0.63   &  \\
\object{HD 150136}  	&  16:41:20.4  & $-$48:45:46.8 & O3.5-4 III(f*) + O6 IV & S14 & 0.43 &  3.27 & 1.64  &  W03  &  --  &  --  & --   &  \\
\object{HD 151804}  	&  16:51:33.7  & $-$41:13:49.9 & O8 Iaf (SB2) & S14 & 0.34	&  4.33 & 1.30  &  V04  &  21.08  &  20.26  & 0.23   &  \\
\object{HD 152248} 	&  16:54:10.1  &  $-$41:49:30.1  & O7 Iabf $+$ O7 Ib(f) & S14 &  0.47  & 3.68   & 1.51   	& V04   & --   & 20.92   &   -- &  \\
\object{HD 152408}  	&  16:54:58.5  &  $-$41:09:03.1&  O8: Iafpe & S14 & 0.44	&  4.17  &  1.75  &  V04  & 21.27  &  20.38  & 0.21  &  \\
\object{HD 152424}  	&  16:55:03.3  &  $-$42:05:27.0 &  OC9.2 Ia (SB1) & S14 & 0.66   & 3.23   & 2.07   & W03   & --   & --   & --   &  \\
\object{HD 153919}  	&  17:03:56.8  &  $-$37:50:38.9 & O6 Iafcp  (SB1E) & S14 &  0.59  & 3.58   &  1.97  &  V04   & --  &--    &  --  &  \\
\object{HD 154043}  	&  17:05:18.9  &  $-$47:04:08.5& B2 Iab & H78 &  0.80  & 3.30  &  2.57  &  W03  &  --  &  --  & --   &  \\
\object{HD 155806}  	&  17:15:19.3  &  $-$33:32:54.3&  O7.5 V((f))z(e) & S14 & 0.27   & 2.48   & 0.57   & W03   &  21.08  &  19.92  &  0.12  &  \\
\object{HD 157246}  	&  17:25:23.7  &  $-$56:22:39.8&  B1 Ib  & H69 & 0.06  &  2.82  &  0.17  &  W03  &  20.68  &  19.23  &  0.07  &  \\
\object{HD 157978}  	&  17:26:19.0  &  $+$07:35:44.3  & G0/K0 II: $+$ A1  &  P98 & 0.64   & 3.72   & 2.42   & W03   &  --  &  --  & --   &  \\
\object{HD 158926}  	&  17:33:36.5  &  $-$37:06:13.8&  B2 IV $+$ DA.79 & H13 & 0.10   & --   &  --  &    &  19.23  &  12.70  & 0.00   &  \\
\object{HD 161056}  	&  17:43:47.0  &  $-$07:04:46.6& B1.5 V & L68	& 0.58  &  --  &  --  &    & 21.23   &  --  & --   &  \\
\object{HD 164073}  	&  18:02:00.6  &  $-$48:48:37.7 &  B3III/IV & H78 & 0.22  &  2.96  &  0.62  &  V04  &  --  &  --  & --   &  \\
\object{HD 164353}	&  18:00:39.0	& $+$02:55:53.6	& B5 Ib	& M50	& 0.10	& --	& --	&  		& 21.00 & 20.26   & 0.27   &  \\
\object{HD 166937}	&  18:13:45.8	& $-$21:03:31.8	& B8 Iab(e)	& G94	& 0.22	& --  &  --  &    & --   &  --  & --   &  \\
\object{HD 167264}  	&  18:15:12.9  &  $-$20:43:41.8 &  O9.7 Iab  & S14 & 0.24  &  3.26  &  0.98  &  V04  & 21.15  &  20.28  &  0.21  &  \\
\object{HD 167838}  	&  18:17:37.7  &  $-$15:25:50.6 & B5 Ia  & M55 &  0.55  & 3.39   & 2.13   & V04   & --   &  --  & --   &  \\
\object{HD 167971}  	&  18:18:05.9  &  $-$12:14:33.3 &  O8 Iaf(n) $+$ O4/5 & S11 & 1.01  &  3.44  &  3.54  &  V04  & 21.6  &  20.85  &  0.26  & \\
\object{HD 169454}  	&  18:25:15.2  &  $-$13:58:42.3 & B1 Ia & W76 & 1.03   &  3.37  & 3.64   & V04   &  --  & --  & --   &  \\
\object{HD 170740}  	&  18:31:25.7  &  $-$10:47:45.0 &  B2 V  & S56 & 0.45  &  3.01  &  1.51  &  V04  &  21.03  &  20.86  &  0.58  & \\
\object{HD 170938}  	&  18:32:37.8  &  $-$15:42:05.9 & B1 Ia  & M55 & 0.97   &  3.34  & 3.47   & W03   & --   &  --  & --   &  \\
\object{HD 171957}  	&  18:38:04.5  &  $-$14:00:16.9 &  B9 IV & C69 &  0.27  &  --    & --    &    &  --  &  --  & --   &  \\
\object{HD 172694}  	&  18:42:16.6  &  $-$16:25:41.0   &  B1: Vne & G68 &  0.34  &  5.04  &  1.86  &  W03  &  --  &  --  & --   &  \\
\object{HD 180554}	&  19:16:13.0  &  $+$21:23:25.4 & B4 IV & L68 & & -- & -- & & -- & -- & -- & \\
\object{HD 184915}  	&  19:36:53.5  &  $-$07:01:38.9 &  B0.5 IIIn & L68	&  0.24  &  3.07  &  0.68  &  W03  &  20.85  &  20.31  &  0.37 &  \\
\object{HD 185418}  	&  19:38:27.5  &  $+$17:15:26.1 &  B0.5 V  & M55 & 0.42   &  2.54  &  1.27  &  V04  & 21.19  &  20.71  &  0.40  &  \\
\object{HD 185859}  	&  19:40:28.3  &  $+$20:28:37.5&  B0.5 Ia  & L68 & 0.56    &  2.74  &  1.64  &  V04  &21.23  &  --  & --   &  \\
\object{HD 186745}  	&  19:45:24.3  &  +23:56:34.4  &  B8 Ia  & M55 &  0.88  &  3.10  &  2.98  &  W03  &  --  &  --  & --   &  \\
\object{HD 186841}  	&  19:45:54.0  	&  +24:05:47.0  &  B0.5 I  & M95 &  0.95  &  2.89  &  3.01  &  W03  &  --  &  --  & --   &  \\
\object{HD 203532}  	&  21:33:54.6  &  $-$82:40:59.1 &  B3 IV  & H69 & 0.30  &  3.37  &  0.94  &  V04  & 20.22  &  20.64  &  0.84  &  \\
\object{HD 303308}  	&  10:45:05.9  &  $-$59:40:06.4  &  O4.5 V((fc)) (SB1?) & S14 & 0.43 &  3.02  &  1.36  &  V04  & 21.41  &  20.34  &  0.15  &  \\ 
\end{longtable}
}	
\end{landscape}

\begin{landscape}
{
\setlength\LTcapwidth{\linewidth}
\begin{longtable}{lllllllllllll}
\\
\caption{Observations planned to complete the EDIBLES sample.\label{tb:targetlist2}}\\
\hline\hline
Identifier 		&  RA  		&  Dec  		&  Spectral type & Ref.  &  \Ebv  &  $R_V$  &  $A_V$  & Ref  	&  log $N$(\ion{H}{i}) &  log $N$(H$_2$)	&  $f_{H_2}$  &  \\
	 			& J2000		& J2000 		&		& &  mag	& 		& mag	& 			 	& cm$^{-2}$	& cm$^{-2}$	& 		&		\\ 
\hline
\endfirsthead
\caption[]{\it{continued}} \\
\hline
Identifier		&  RA  		&  Dec  		&  Spectral type & Ref.  &  \Ebv  &  $R_V$  &  $A_V$  & Ref  	&  log N(\ion{H}{i}) &  log N(H$_2$)	&  $f_{H_2}$  &  \\
	 			& J2000		& J2000 		&		& &  mag	& 		& mag	& 			 	& cm$^{-2}$	& cm$^{-2}$	& 		&		\\ 
\endhead
\hline
\multicolumn{13}{r}{\it{continued on next page}}\\
\endfoot
\hline
\endlastfoot
\object{HD 36982}  		&  05:35:09.8  &  $-$05:27:53.3  &  B1.5 Vp  &  S52	& 0.31  & 4.84  &  1.74  &  V04  & 21.57   & --   & --   &  \\ 
\object{HD 37020}  		&  05:35:15.8  &  $-$05:23:14.3  &  B0.5 V  & L76 & 0.28  &  6.33  &  1.77  &  V04  & 21.48   & --   & --   &  \\ 
\object{HD 37021}  		&  05:35:16.1  &  $-$05:23:07.3  &  B0 V  & G68 & 0.30 &  5.84  &  2.80  &  V04  &  21.68  &  --  & --   &  \\
\object{HD 37023}  		&  05:35:17.2  &  $-$05:23:15.6  &  B1.5 Vp  & L06 & 0.37 &  4.82	& 1.74 & V04 &  --  &  --  & --   &  \\
\object{HD 37128}  		&  05:36:12.8  &  $-$01:12:06.9  &  B0 Ia  & M50 & 0.09  &  --  &  --  &    &  20.45  &  16.57  &  0.04  &  \\
\object{HD 38087}  		&  05:43:00.6  &  $-$02:18:45.4  &  B5 V  & S71 & 0.28 &  4.93  &  1.48  &  V04  &  20.91  &  20.64  & 0.52   &  \\
\object{HD 39680} 		&  05:54:44.7  &  $+$13:51:16.9  &  O6 V:[n]pe var & S11 &  0.33  &  4.73  &  1.51  &  W03  &  21.3  &  19.53  & 0.03   &  \\
\object{HD 45314}  		&  06:27:15.8  &  $+$14:53:21.1  &  O9: npe  & S11 & 0.13 &  4.42  &  1.90  &  V04  & 21.04  &  20.60  &  0.42  &  \\
\object{HD 49787}  		&  06:49:55.5  &  $-$05:30:47.5  &  B1 V:pe & M55 &  0.20  &  2.39  &  0.43  &  W03  &  --  &  --  & --   &  \\
\object{HD 50820}  		&  06:54:42.0  &  $-$01:45:23.4  &  B3 IVe $+$ K2 II  & H82 &  0.72  &  4.39  &  3.38  &  W03  &  --  &  --  &  --  &  \\
\object{HD 61827}  		&  07:39:49.3  &  $-$32:34:42.2  &  B3 Iab  & G77 &  0.78  &  2.98  &  2.68  &  W03  &  --  &  --  &    &  \\
\object{HD 114886}  		&  13:14:44.4  &  $-$63:34:51.8  &  O9 III $+$ O9.5 III  & S14 &  0.28  &  2.90  &  0.84  &  W03  &  21.34  &  20.23  & 0.13  &  \\
\object{HD 156201}  		&  17:17:45.5  &  $-$35:13:27.0  & B0.5 Ia  & G77 & 0.91   & 3.23   & 2.78   & W03   &  --  &  --  & --   &  \\
\object{HD 164740}  		&  18:03:40.0  	&  $-$24:22:43.0  &  O7: V $+$ sec. & S14 & 0.90   &  5.03  &  4.48  &  V04  & 21.95  &  20.19  & 0.03   &  \\
\object{HD 164906}  		&  18:04:25.8  &  $-$24:23:08.3& B0 Ve	& L06 & 0.44   & 4.57   & 2.06   &  V04  &  -- & --   & --   &  \\
\object{HD 165319}  		&  18:05:58.8  &  $-$14:11:53.0& O9.7 Ib & S11 & 0.78   & 3.25   & 2.70   &  W03  &  -- & --   & --   &  \\
\object{HD 168076}  		&  18:18:36.4  &  $-$13:48:02.0&  O4 III(f) & S11  & 0.76   &  3.47  &  2.64  &  V04  & 21.65   & 20.68 & 0.18   &  \\
\object{HD 183143}  		&  19:27:26.6  &  +18:17:45.2    & B7 Ia  & M55 & 1.19   & --   & --   &    &    --  &  --  & -- & \\
\end{longtable}
}	
\end{landscape}

\clearpage

\section{Spectrum of HD170740}\label{appendix:fullspec}

\begin{figure}[ht!]
\centering
\includegraphics[viewport=10 25 575 825,width=15.25cm]{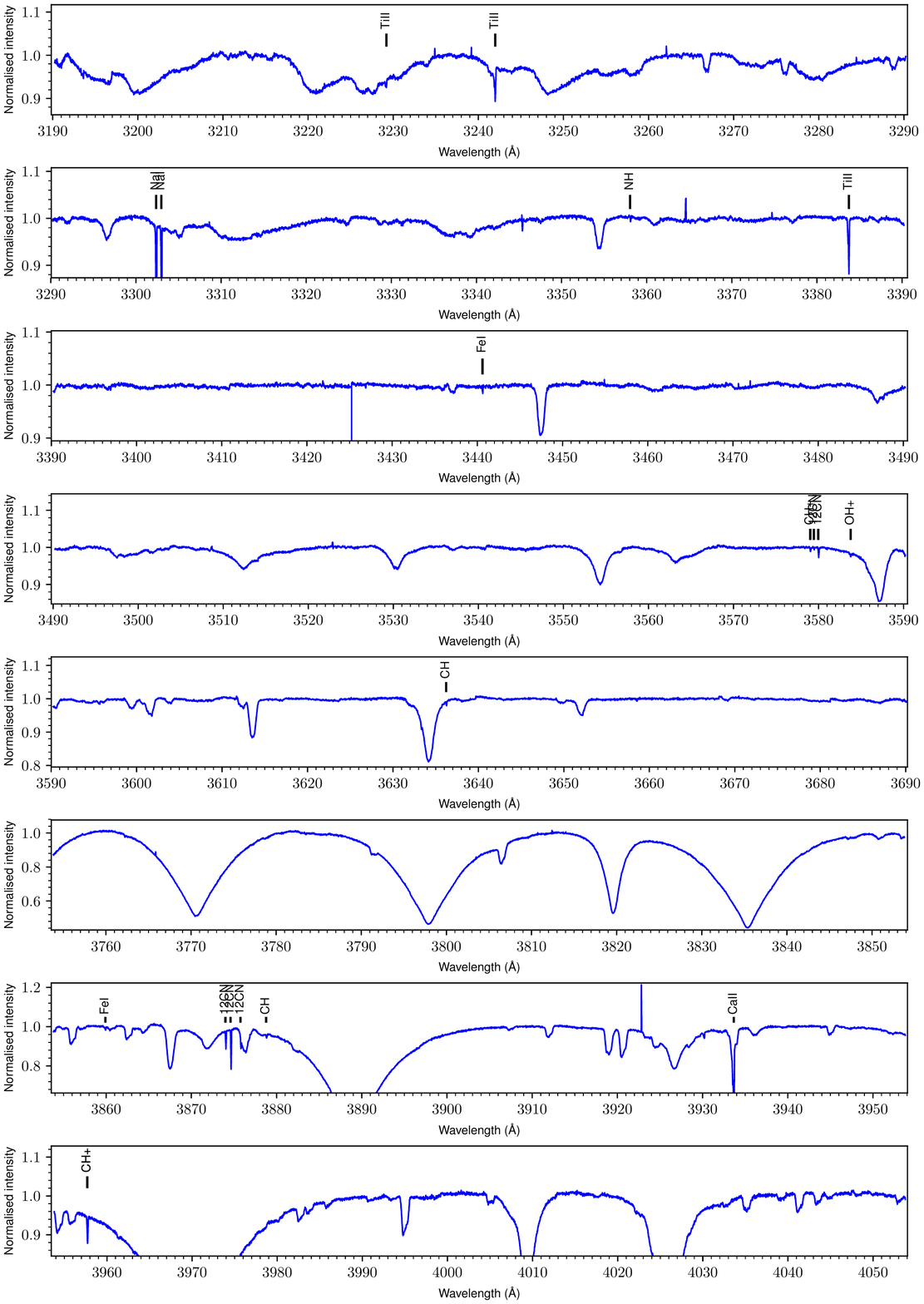}
\caption{EDIBLES spectrum for HD\,170740 (solid blue).
The preliminary normalisation shown here is from spline fits to broad regions of each UVES settings, and order-merging residuals can also be seen -- future quantitative analysis will employ tailored, local rectifications of the data. 
Electronic transitions of interstellar atomic and di-atomic species are labeled. Above 4000~\AA\ a generic model of the telluric transmission spectrum (retrieved from the ESO Sky Calculator) is shown in yellow/orange as well as the \citet{1994A&AS..106...39J} average ISM DIB spectrum (scaled to \Ebv~= 0.5 mag) in solid green.}\label{fig:appendix:fullspec}
\end{figure}

\addtocounter{figure}{-1}

\begin{figure}
\centering
\includegraphics[viewport=10 25 575 825,width=17.cm]{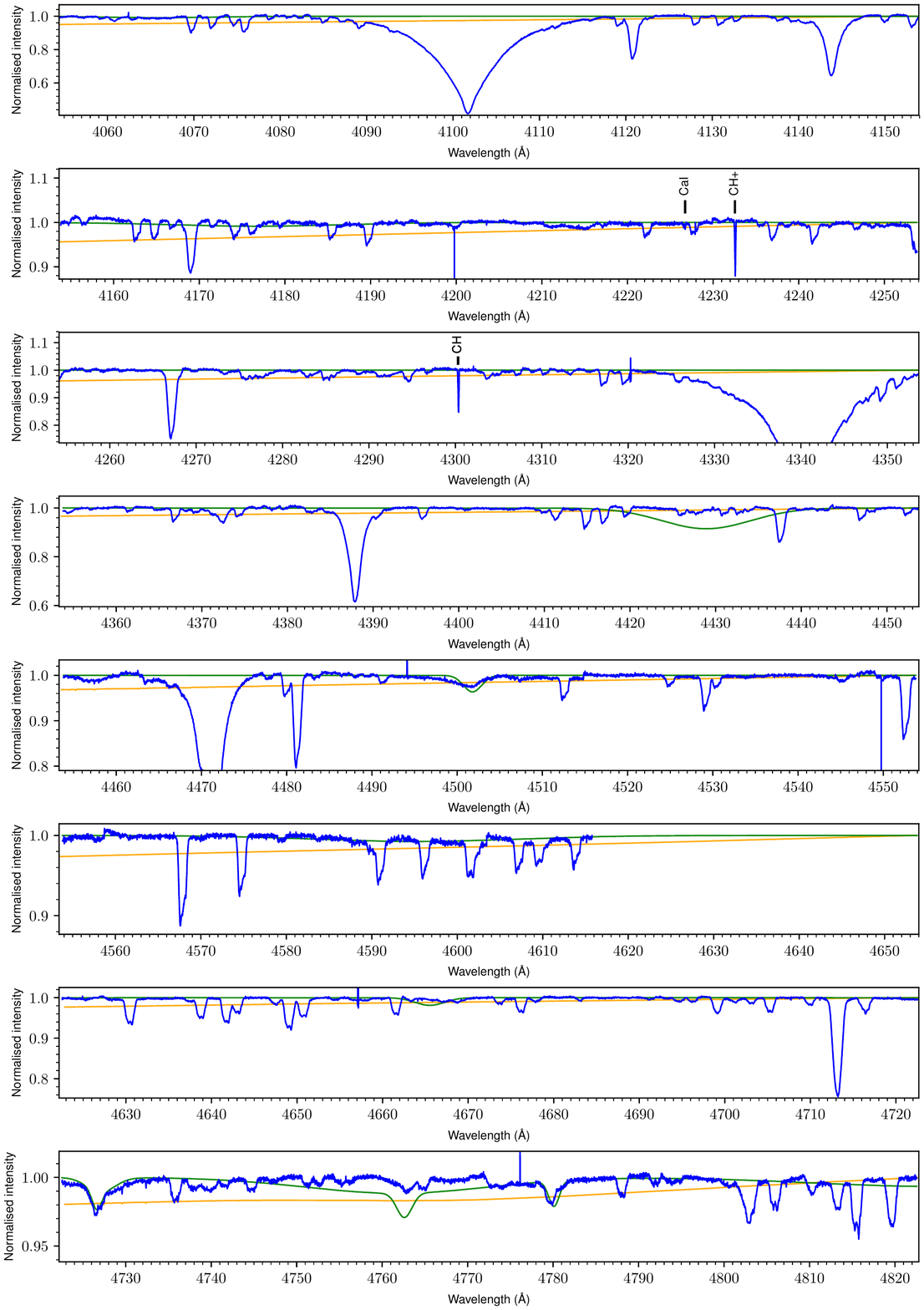}
\caption{EDIBLES spectrum for HD\,170740 (continued).}
\end{figure}

\addtocounter{figure}{-1}

\begin{figure}
\centering
\includegraphics[viewport=10 25 575 825,width=17.cm]{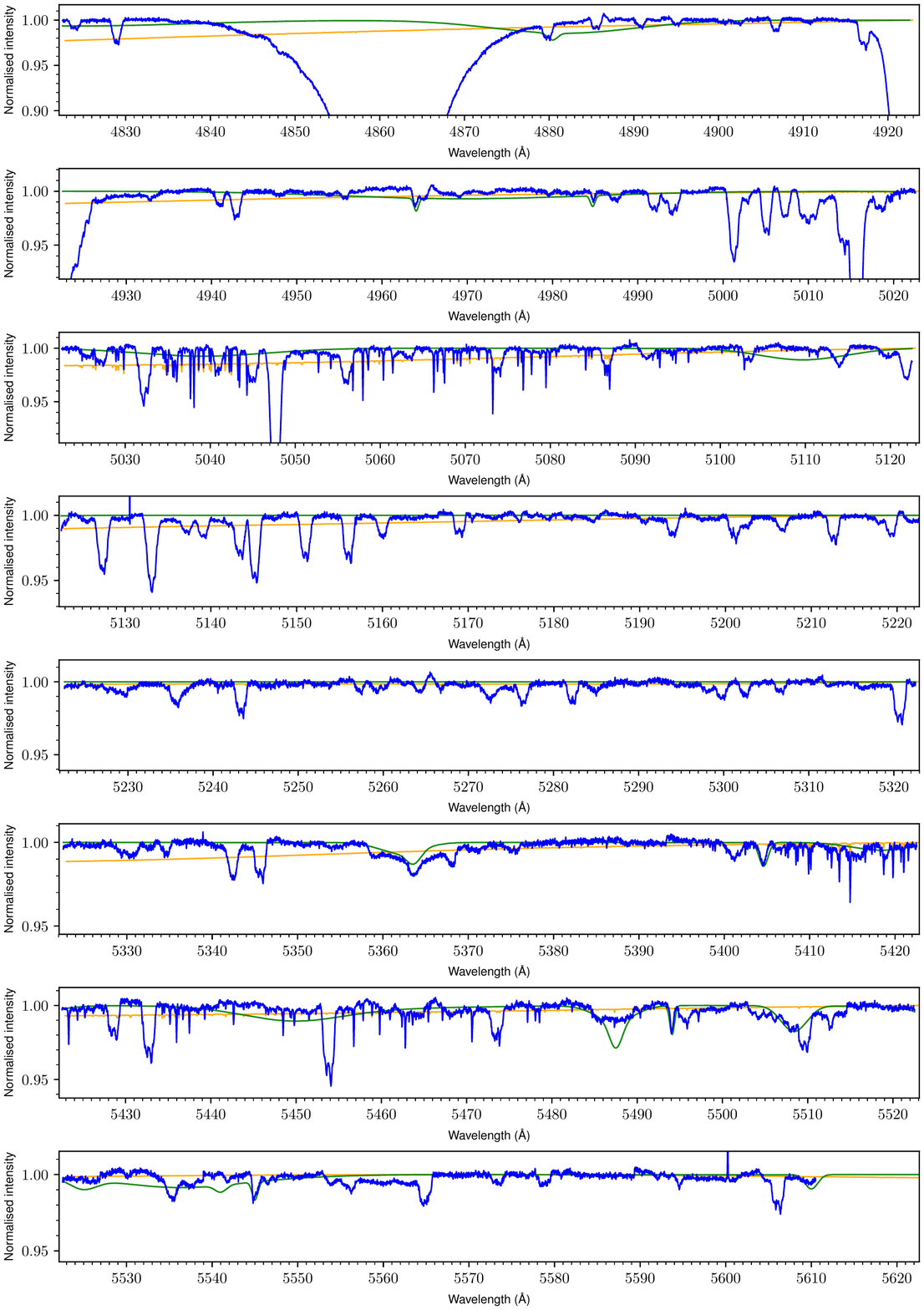}
\caption{EDIBLES spectrum for HD\,170740 (continued).}
\end{figure}
\addtocounter{figure}{-1}

\begin{figure}
\centering
\includegraphics[viewport=10 25 575 825,width=17cm]{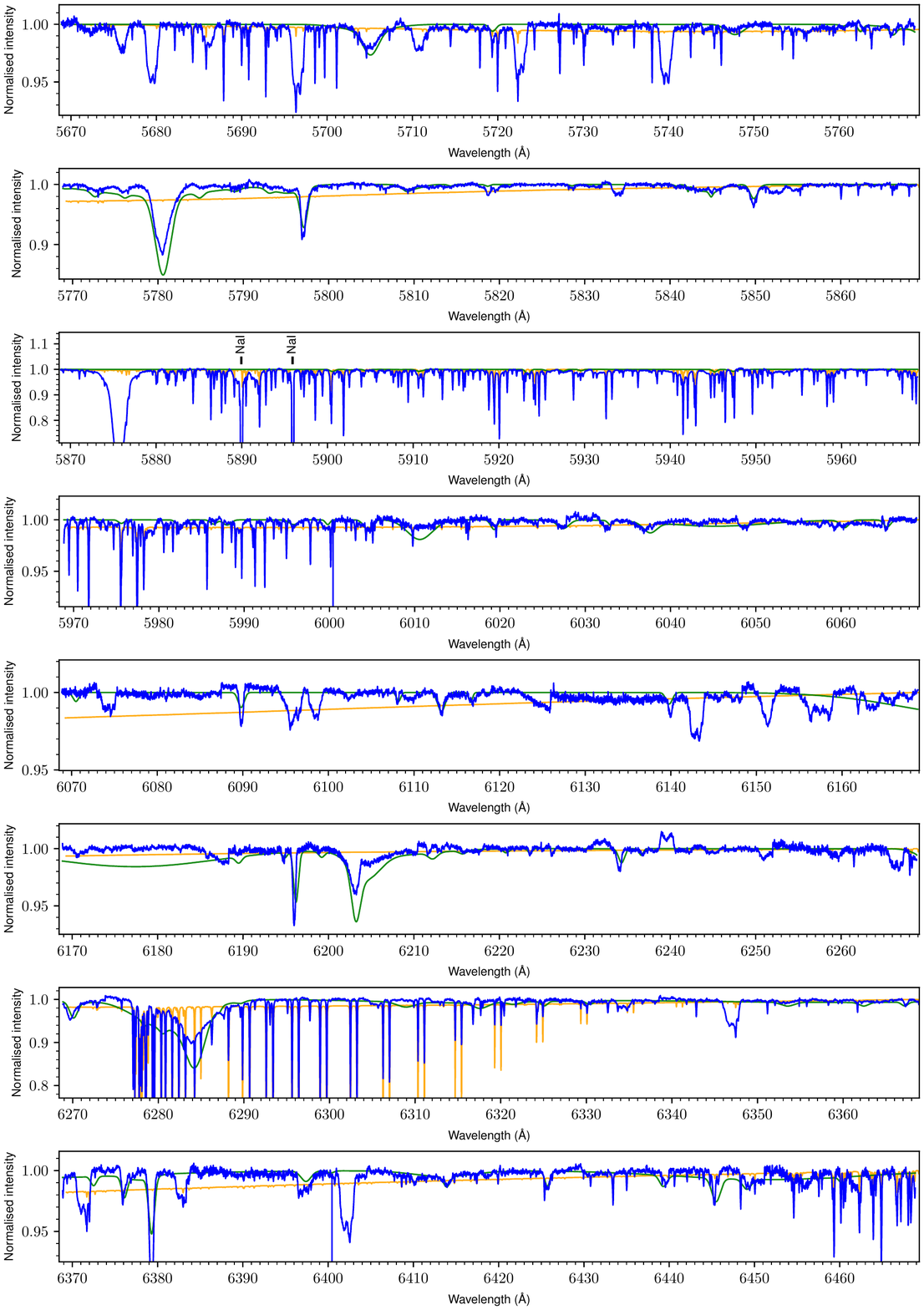}
\caption{EDIBLES spectrum for HD\,170740 (continued).}
\end{figure}
\addtocounter{figure}{-1}

\begin{figure}
\centering
\includegraphics[viewport=10 25 575 825,width=17cm]{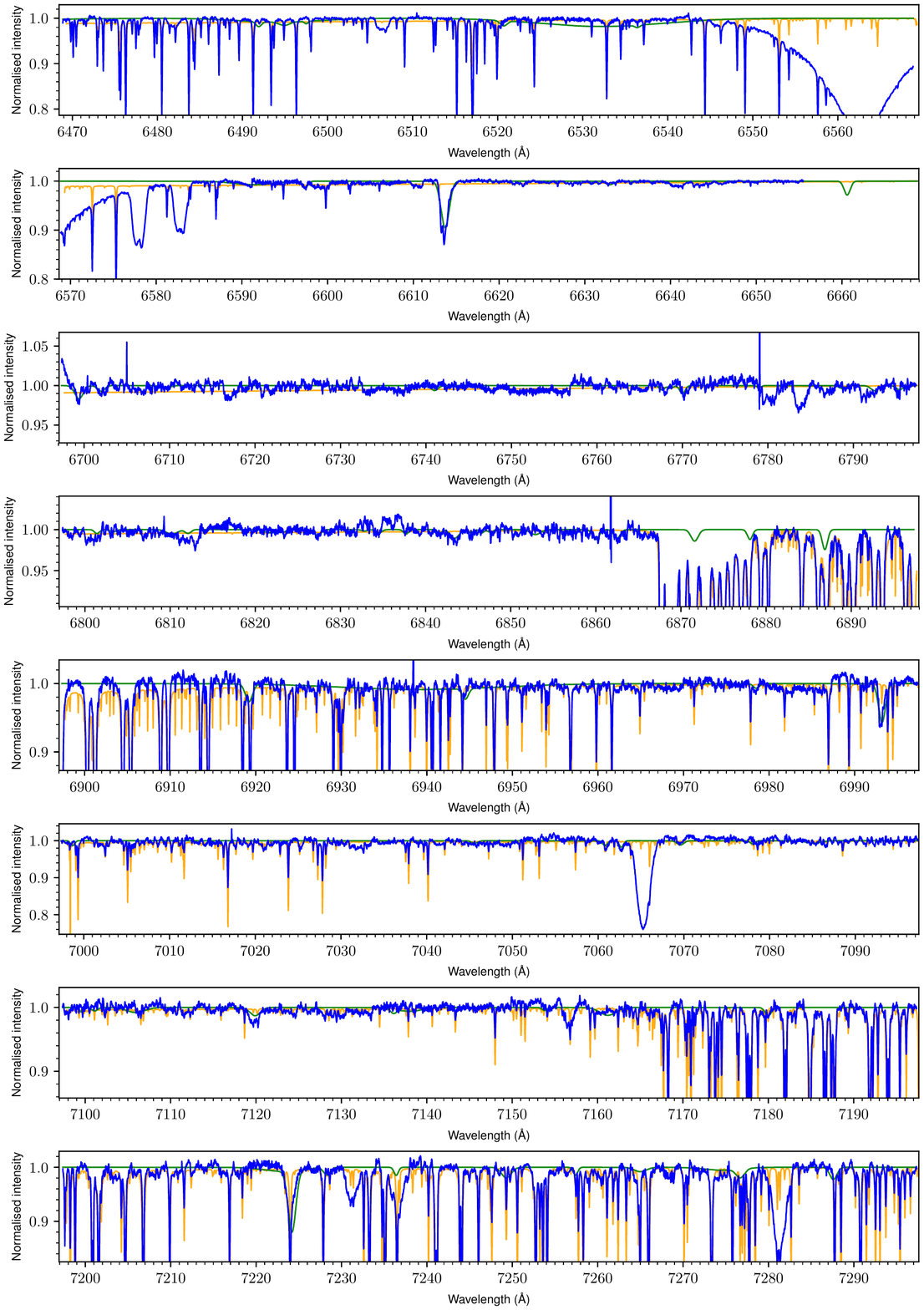}
\caption{EDIBLES spectrum for HD\,170740 (continued).}
\end{figure}
\addtocounter{figure}{-1}

\begin{figure}
\centering
\includegraphics[viewport=10 25 575 825,width=17cm]{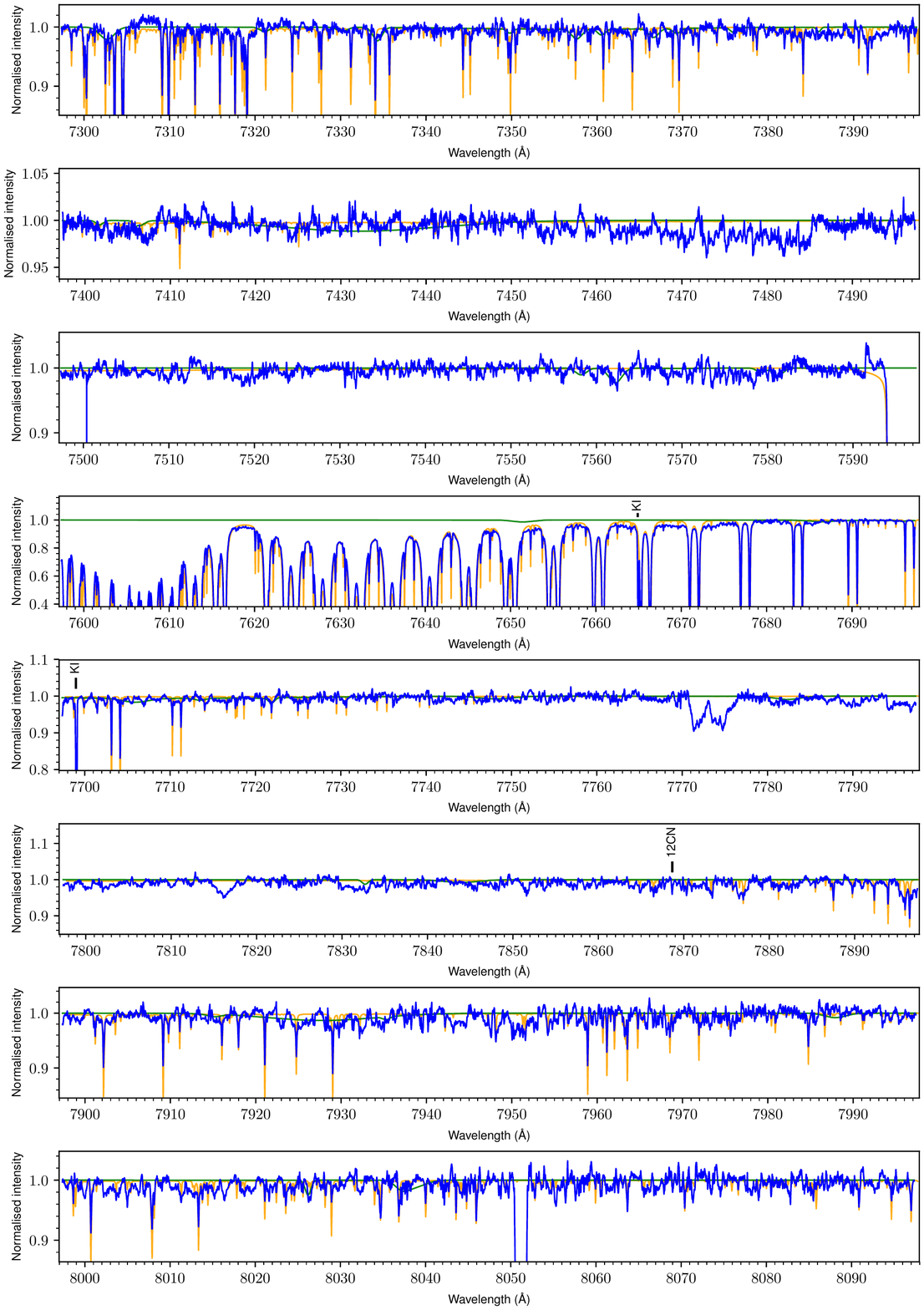}
\caption{EDIBLES spectrum for HD\,170740 (continued).}
\end{figure}
\addtocounter{figure}{-1}

\begin{figure}
\centering
\includegraphics[viewport=10 25 575 825,width=17cm]{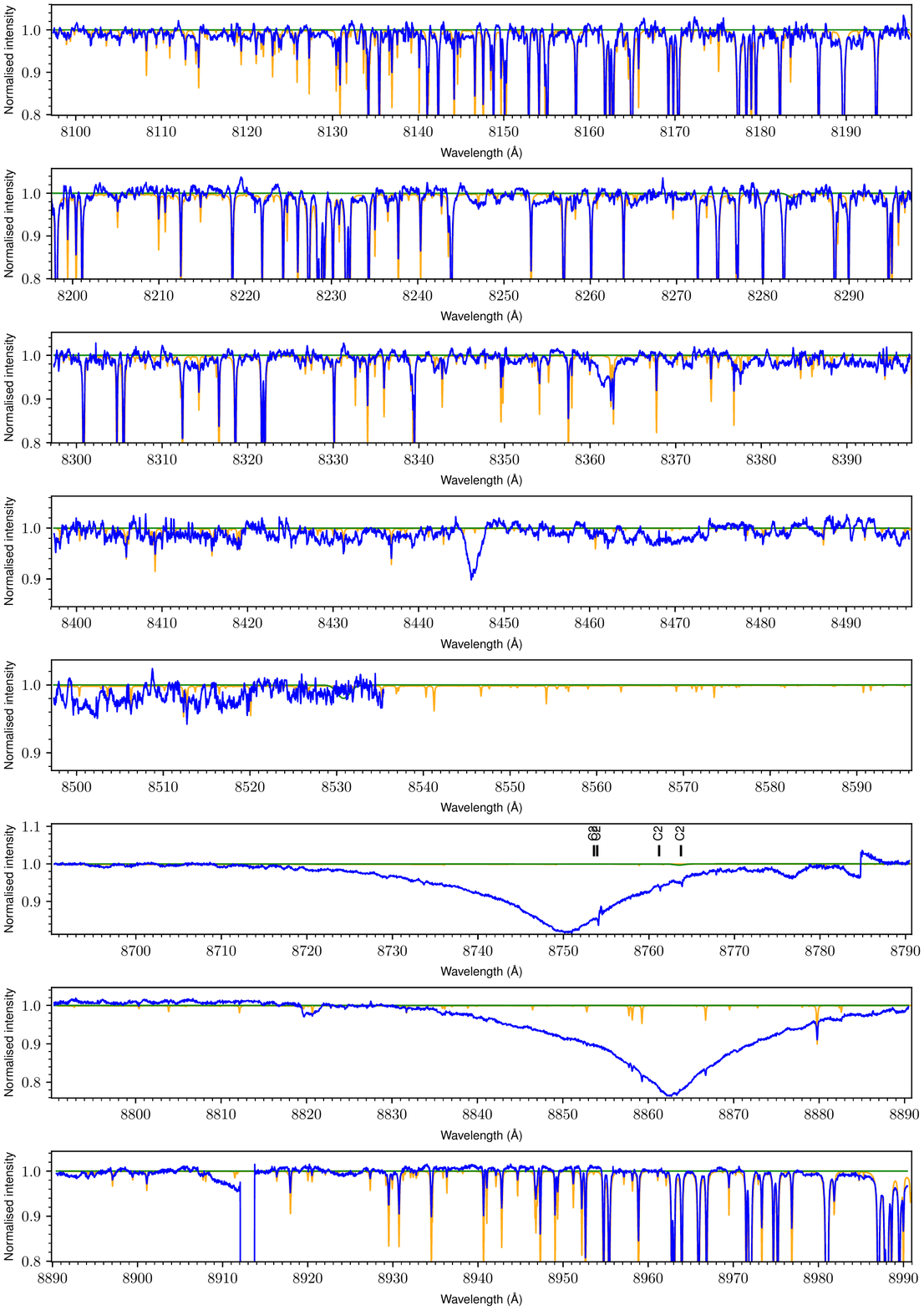}
\caption{EDIBLES spectrum for HD\,170740 (continued).}
\end{figure}
\addtocounter{figure}{-1}

\begin{figure}
\centering
\includegraphics[viewport=10 25 575 825,width=17cm]{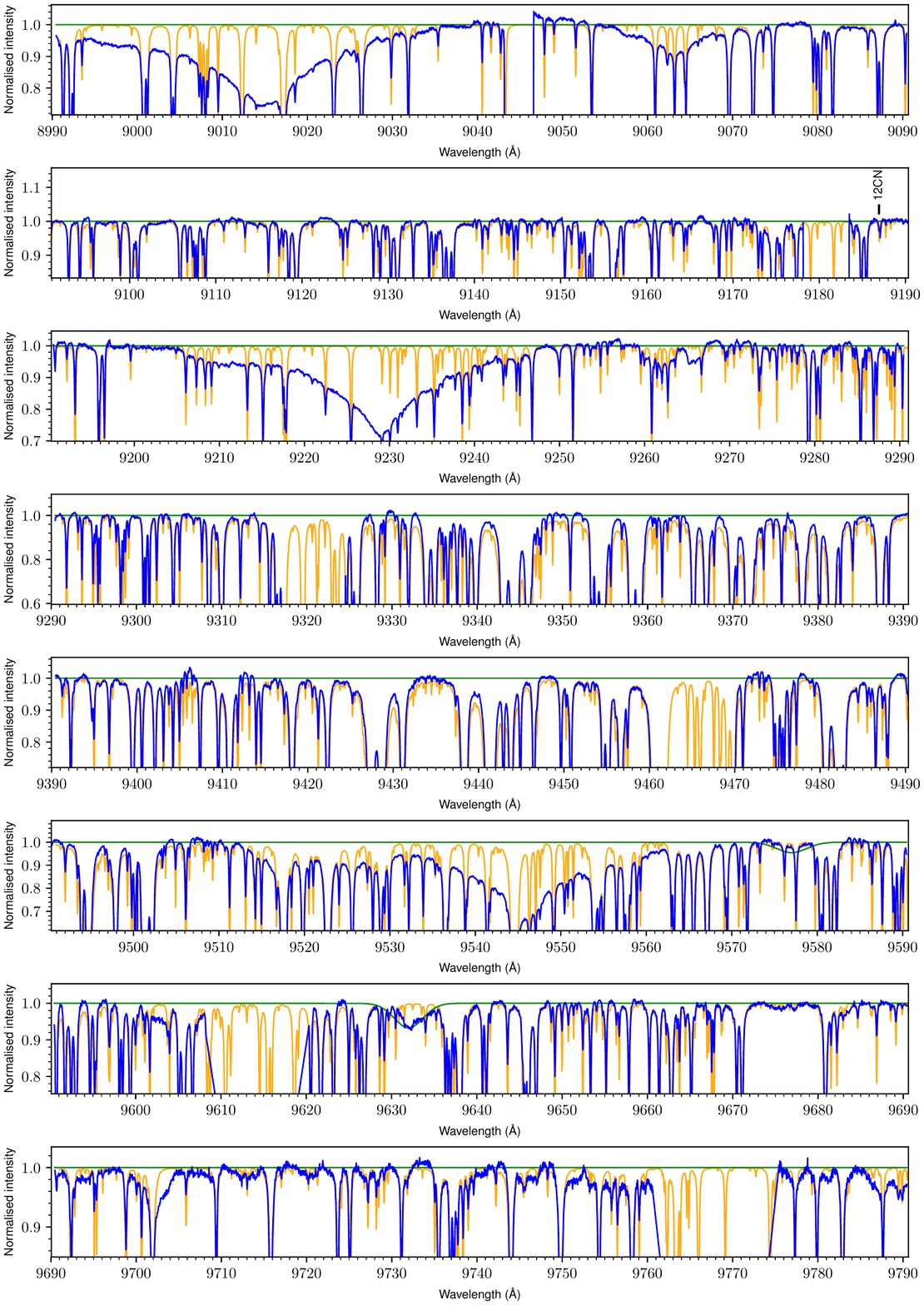}
\caption{EDIBLES spectrum for HD\,170740 (continued).}
\end{figure}
\addtocounter{figure}{-1}

\begin{figure}
\centering
\includegraphics[viewport=10 525 575 825,width=17cm]{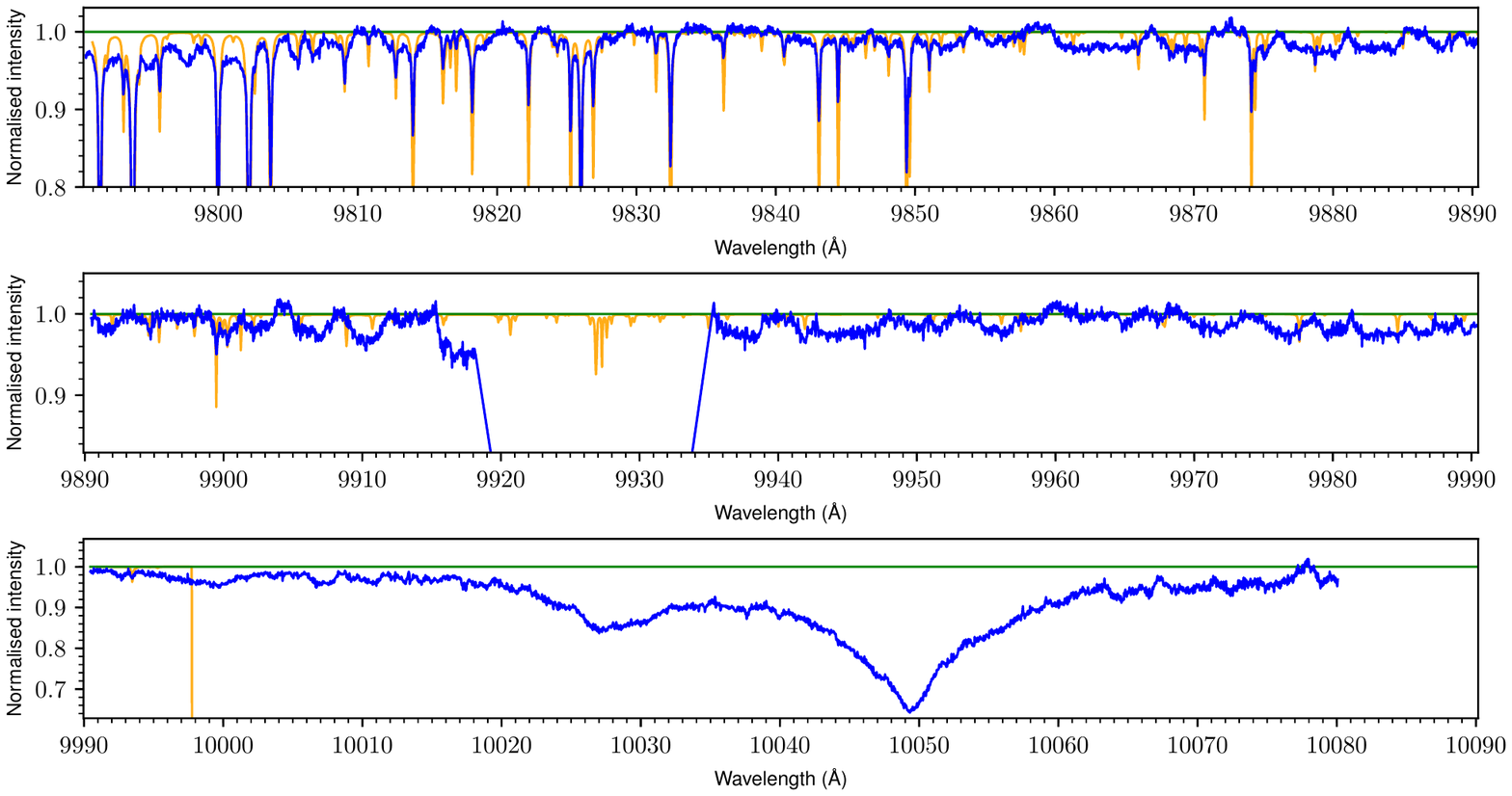}
\includegraphics[viewport=10 525 575 825,width=17cm]{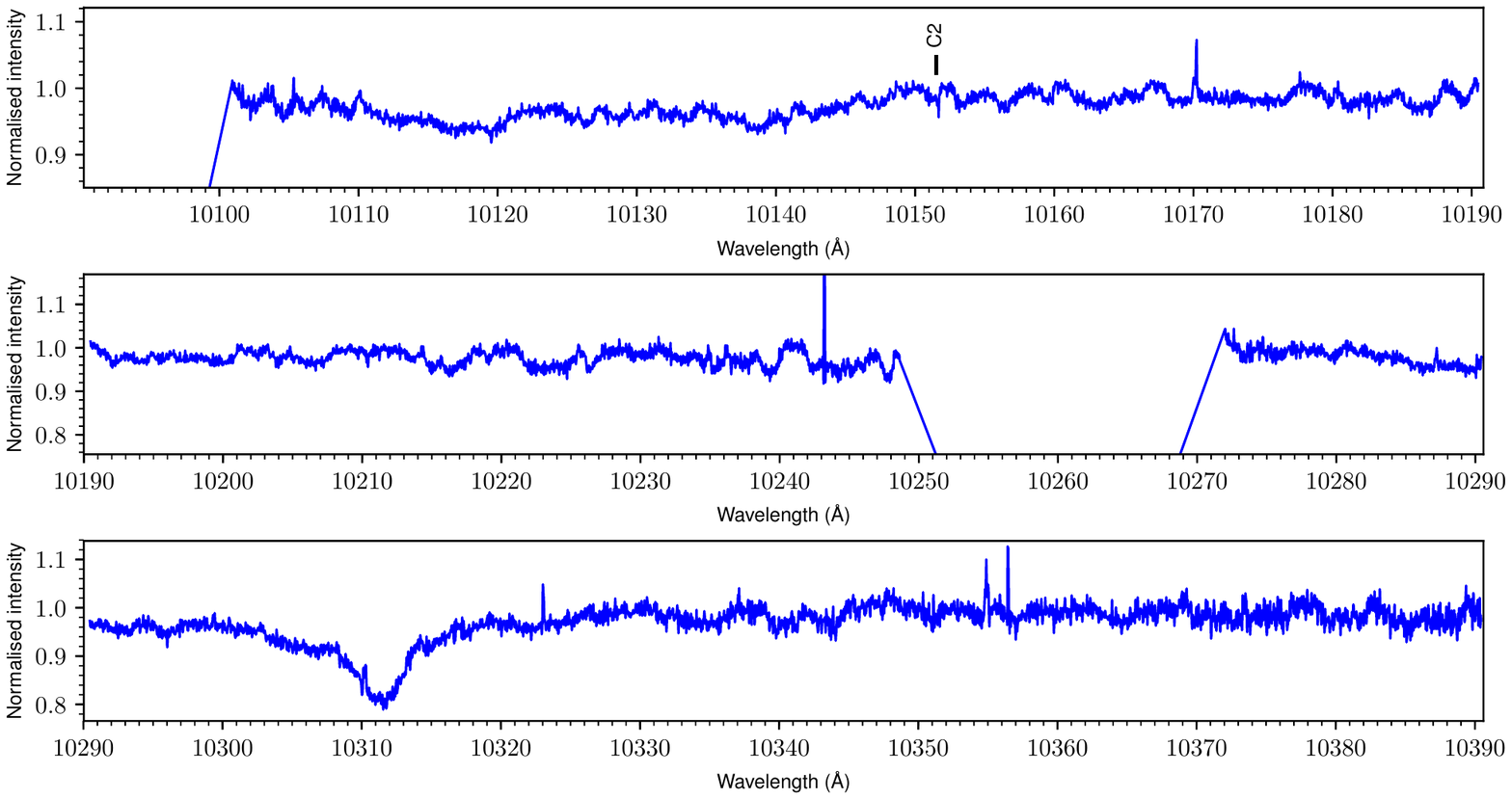}
\caption{EDIBLES spectrum for HD\,170740 (continued).}
\end{figure}
\addtocounter{figure}{-1}

\label{lastpage}

\end{document}